\documentclass[12pt]{IOPART}
\usepackage{graphicx}
\def\lbets{$\lambda$-(BETS)$_{2}$GaCl$_{4}$}
\def\hc2{$H_{c2}$}
\def\betasf{$\beta ''$-(BEDT-TTF)$_2$SF$_5$CH$_2$CF$_2$SO$_3$}
\def\cuscn{$\kappa$-(BEDT-TTF)$_2$Cu(NCS)$_2$}
\def\khg{$\alpha$-(BEDT-TTF)$_2$KHg(SCN)$_4$}
\def\nh4{$\alpha$-(BEDT-TTF)$_2$NH$_4$Hg(SCN)$_4$}

\begin{document}
\setcounter{footnote}{1}

\title[Quasi-two-dimensional organic superconductors]
{Quasi-two-dimensional organic superconductors: a review.}
\author{John Singleton$^{1,2}$ and Charles Mielke.}

\address{$^1$National High Magnetic Field Laboratory,
Los Alamos National Laboratory, MS-E536,
Los Alamos, New Mexico 87545, USA}

\address{$^2$University of Oxford, Department of Physics, The
Clarendon Laboratory, Parks Road,
Oxford OX1 3PU, U.K.}

\begin{abstract}
We present a review of quasi-two-dimensional
organic superconductors.
These systems exhibit many interesting phenomena, including
reduced dimensionality, strong electron-electron
and electron-phonon interactions and the proximity of antiferromagnetism,
insulator states and superconductivity.
Moreover, it has been possible to measure the
electronic bands of many of the organics
in great detail, in contrast to the situation in
other well-known systems in which similar phenomena occur.
We describe the crystal structure and normal-state
properties of the organics,
before presenting the experimental evidence
for and against exotic superconductivity mediated by
antiferromagnetic fluctuations.
Finally, three instances of field-induced
unconventional superconductivity will be described.
\end{abstract}

\maketitle

\section{Introduction}
Over the past fifteen years there has been
an increasing scientific interest in crystalline organic
metals and superconductors.
These materials are made up of relatively
small organic molecules constructed from carbon,
sulphur, selenium, hydrogen and the like,
combined with a second molecular species,
either inorganic or organic.
One usually thinks of solids made up
from organic molecules as insulators;
however, the crystalline organic
metals conduct electricity,
and their low temperature resistivity
usually decreases with decreasing temperature,
a classic signature of more conventional (elemental) metals.

The reason for the current
intense interest in organic
metals is that they form a uniquely flexible system for
the study of superconductivity, magnetism
and many-body effects~\cite{review,mck1,kanoda,aoki,schmalian}.
In this context, the term ``many-body effects''
refers to the interactions experienced by the mobile
electrons within the organic metal;
in such systems, the electrons can no longer
be thought of as independent entities,
but instead interact with each other and
with excitations such as phonons (i.e.
the vibrations of the crystal lattice).
Materials in which such many-body effects
are important are often referred to as
{\it correlated electron systems}.

Of course there are other
interesting correlated-electron system.
Prime examples are the so-called ``High $T_{\rm c}$''
cuprate superconductors (e.g. YBa$_2$Cu$_3$O$_7$)
which have been the subject of considerable
experimental and theoretical effort.
Organic metals share many features with
the cuprates-- a quasi-two-dimensional bandstructure, the
proximity of antiferromagnetism, superconductivity and
insulating behaviour, unconventional
superconductivity~\cite{review,mck1,kanoda,aoki,schmalian}--
and yet they are much cleaner systems;
it has been possible to measure many
details of the electronic energy levels
of the organics using accurate
techniques~\cite{review,wosnitza}, whereas this has
proved impossible in the cuprates.
Finally, the organics are interesting because a variety
of exotic phases have been predicted and
measured at potentially accessible
magnetic fields ($B \sim 50-100$~T)
and temperatures~\cite{uji,janeloff,frohlich}.

The description ``organic superconductor''
can encompass quite a range of materials
(e.g. quasi-one-dimensional
superconductors of the form
(TMTSF)$_2$X~\cite{chaikreview}, derivatives
of buckminsterfullerene~\cite{ishiguro}
and the recently-discovered field-effect
devices based on acene and its relatives~\cite{bell}).
Each of these is worth a review article in its own right.
To avoid information overload,
we shall therefore concentrate solely on the quasi-two-dimensional
organic superconductors, which form a rich and diverse family of
materials~\cite{review,ishiguro}.
These systems are also of interest because they are
perhaps the closest organic relative of the
above-mentioned cuprate superconductors~\cite{mck1,kanoda}.

The review will describe the normal state properties of the organics,
followed by a discussion of the proposed mechanisms for superconductivity.
Afterwards, the superconducting phase diagram will be reviewed.
Finally, we shall mention some recent instances of (in some cases exotic)
field-induced superconductivity.

\section{Normal-state properties}
\subsection{Crystal structure.}
\label{s2p1}
\begin{figure}[htbp]
\centering
\includegraphics[height=0.75in]{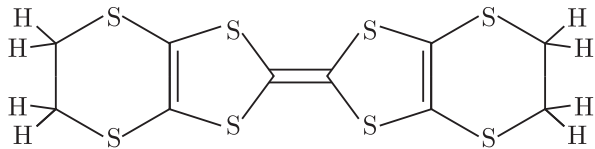}
\caption{The bisethylenedithiotetrathiofulvalene (BEDT-TTF) molecule.
}
\label{etmolecule}
\end{figure}

The electronic bands\footnote{The term {\it band}
refers to the distribution of mobile electronic states
within a solid. As we shall see below,
bands are characterised by
{\it dispersion relationships},
formulae which relate the energies $E({\bf k})$
of the mobile
electronic states to their wavevector quantum
numbers {\bf k}. For a summary, see e.g.
Chapters 2-4 of Reference~\cite{singlebook}.}
of quasi-two-dimensional organic superconductors
are derived from {\it molecular} orbitals\cite{mori,ishiguro}.
A number of molecular species are used as bandstructure building blocks;
bis(ethylenedithio)tetrathiofulvalene (also known as BEDT-TTF or ET;
Figure~\ref{etmolecule}) is a typical example~\cite{mori,ishiguro,montgomery}.

Although such molecules look complicated
(and their acronyms may seem daunting)
at first sight,
the principles behind their use in
the construction of crystalline
organic metals are very simple.
For example, the BEDT-TTF molecule
is roughly flat, so that it can be packed in a variety
of arrangements in a solid, and it is surrounded by
voluminous molecular orbitals;
to create electronic bands, it is merely necessary
to stack the BEDT-TTF molecules next to each other,
so that the molecular orbitals can overlap.
Crudely one might say that
this enables the electrons to transfer from molecule to molecule.

Most of the molecules used to create quasi-two-dimensional organic superconductors
are of a similar form to BEDT-TTF; for example, when
the innermost four sulphur atoms of BEDT-TTF are replaced by
selenium, one obtains BETS (bis(ethylenedithio)tetraselenafulvalene)~\cite{mori},
which can also be used to produce organic superconductors.
Other varieties include BEDO~\cite{mori,bedo} and MDT-TTF~\cite{mori}.

Returning to our initial example,
in order to form the bandstructure,
BEDT-TTF molecules must be
stacked next to each other so that the
molecular orbitals overlap.
This arrangement of the bandstructure-forming molecules
in an ordered arrangement is usually accomplished
by making a {\it charge-transfer salt}~\cite{review,mori,ishiguro,montgomery}.
In a charge-transfer salt, a number $j$ of BEDT-TTF
molecules will jointly donate an electron to a
second type of molecule (or collection of molecules)
which we label $X$, to form
the compound (BEDT-TTF)$_jX$;
owing to its negative charge, $X$ is
known as the {\it anion},
while the BEDT-TTF molecule is sometimes referred to as
the {\it donor} or {\it cation}.
The transfer of charge serves to bind the
charge-transfer salt together (in a manner
analogous to ionic bonding) and also leaves behind
a hole,\footnote{A hole is an empty electronic
state in an electronic band.
As far as electrical conduction
is concerned, the properties of a hole are formally
equivalent to those of a {\it positively}
charged particle with the same velocity
as that of the (empty) electronic state.
For a simple introduction to
the concept of holes, see e.g. Reference~\cite{singlebook},
Chapter~5.}
jointly shared between the $j$ BEDT-TTF
molecules.
This means that the bands formed by the
overlap of the BEDT-TTF molecular orbitals will be
{\it partially filled}, leading one to expect that
the charge-transfer salt will conduct electricity.
\begin{figure}[htbp]
\centering
\includegraphics[height=5cm]{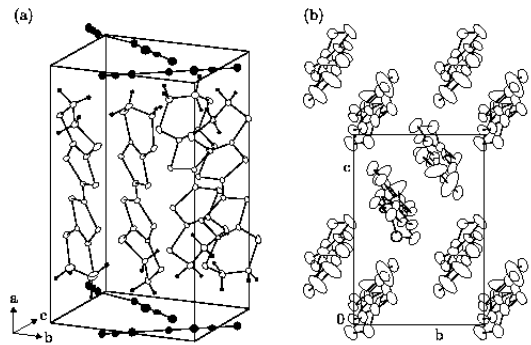}
\caption{Structure of the BEDT-TTF
charge-transfer salt $\kappa$-(BEDT-TTF)$_2$Cu(NCS)$_2$
(a $T_{\rm c} \approx 10.4$~K superconductor).
(a) Side view of molecular arrangement; the
BEDT-TTF molecules pack in planes separated
by layers of the smaller Cu(NCS)$_2$ anions.
(b) View downwards onto the BEDT-TTF planes,
showing that the BEDT-TTF molecules are packed
closely togther, allowing substantial overlap
of the molecular orbitals.
The unit cell edges are
16.248~\AA, 8.44~\AA ~and 13.124~\AA ~at room temperature.
(After Reference~\cite{crystal}.)}
\label{kappapic}
\end{figure}

Figure~\ref{kappapic}~\cite{crystal} shows the molecular arrangements
in the BEDT-TTF
charge-transfer salt $\kappa$-(BEDT-TTF)$_2$Cu(NCS)$_2$,
a $T_{\rm c} \approx 10.4$~K superconductor.
(Here the $\kappa$ denotes the packing arrangement
of the BEDT-TTF molecules; often several different
packing arrangements can be achieved with one particular
anion.)
The BEDT-TTF molecules are packed into layers,
separated by layers of the Cu(NCS)$_2$ anion molecules.
Within the BEDT-TTF layers, the molecules are in
close proximity to each other, allowing substantial
overlap of the molecular orbitals;
the transfer integrals~\cite{singlebook},
which parameterise the ease of
hopping of electrons between BEDT-TTF molecules, will
be relatively large within the BEDT-TTF planes.
Conversely, in the direction perpendicular to the
BEDT-TTF planes, the BEDT-TTF molecules are well
separated from each other; the
transfer integrals will be much smaller
in this direction (i.e. hopping is more difficult).
This extreme anisotropy, resulting from a layered
structure, is typical of most quasi-two-dimensional organic
superconductors; it results in
electronic properties which for many purposes can be considered
to be two dimensional.
We shall return to the question of dimensionality below.
\begin{figure}[htbp]
\centering
\includegraphics[height=5cm]{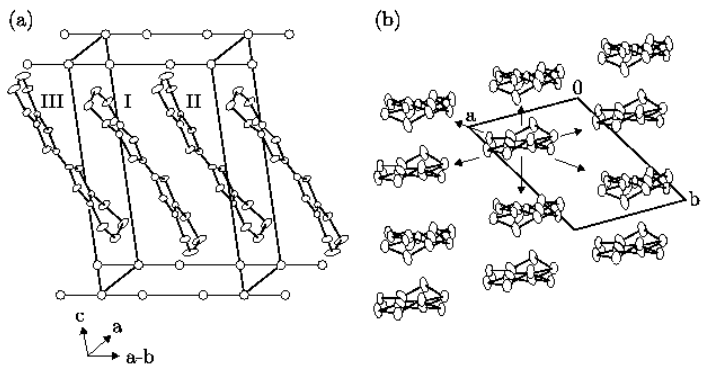}
\caption{Structure of the BEDT-TTF
charge-transfer salt $\beta$-(BEDT-TTF)$_2$I$_3$.
(a)~Side view of molecular arrangement.
As in Figure~\ref{kappapic}, the
BEDT-TTF molecules pack in planes separated
by layers of the smaller anions; in this case the
anions are I$_3$ molecules.
(b)~View downwards onto the BEDT-TTF planes,
showing that the BEDT-TTF molecules are packed
closely togther, allowing substantial overlap
of the molecular orbitals~\cite{triiodide}.
However, the packing arrangement of the BEDT-TTF
molecules is considerably different from that
in $\kappa$-(BEDT-TTF)$_2$Cu(NCS)$_2$
(see Figure~\ref{kappapic}).}
\label{beta}
\end{figure}

Figure~\ref{beta} shows another BEDT-TTF charge-transfer salt,
$\beta$-(BEDT-TTF)$_2$I$_3$ (again $\beta$ denotes the
packing arrangement of the BEDT-TTF molecules)~\cite{triiodide}.
As in the case of $\kappa$-(BEDT-TTF)$_2$Cu(NCS)$_2$
(Figure~\ref{kappapic}), the
BEDT-TTF molecules pack in planes separated
by layers of the smaller anions, in this case
I$_3$ molecules.
However, the packing arrangement of the BEDT-TTF
molecules is considerably different from that
in $\kappa$-(BEDT-TTF)$_2$Cu(NCS)$_2$.

A comparison of Figures~\ref{kappapic} and \ref{beta}
shows that charge-transfer salts are an extremely
flexible system with which to study the physics of band
formation; by changing the anion, one can make the
BEDT-TTF molecules pack in different arrangements.
As the BEDT-TTF molecules are long and flat (unlike
single atoms),
the transfer integrals will depend
strongly on the way in which the BEDT-TTF molecules
are arranged with respect to each other;
this will be reflected in the {\it shape} of the resulting bands~\cite{mori,ishiguro}.
It is therefore convenient to classify
charge-transfer salts according to their
crystal structures, as each packing arrangement results
in a distinct electronic-band topology.
The main arrangements are labelled the $\alpha$,
$\beta$, $\beta''$, $\delta$,
$\chi$, $\kappa$, $\theta$ and $\lambda$ phases~\cite{mori}.

\begin{figure}[htbp]
\centering
\includegraphics[height=6cm]{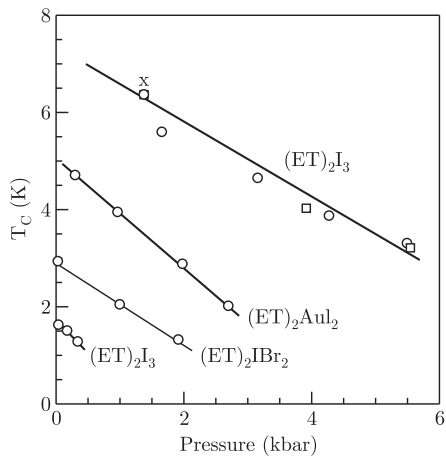}
\caption{Superconducting
critical temperature $T_{\rm c}$ versus hydrostatic pressure
for three $\beta$-phase BEDT-TTF (ET) salts.
Decreasing the unit cell size, either by using a shorter
anion or by increasing the pressure, reduces $T_{\rm c}$~\cite{betapressure}.
(Note that $\beta$-(BEDT-TTF)$_2$I$_3$ undergoes a structural
phase transition at about 0.6~kbar.)
}
\label{unitcell}
\end{figure}

More subtle changes are possible {\it within} a particular morphology.
It is possible to make $\beta$-phase salts
of the form $\beta$-(BEDT-TTF)$_2X$ using several different
anion molecules $X$, for example I$_3$, IBr$_2$, AuI$_2$ {\it etc.}.
The different possible anions have slightly different lengths;
therefore the size of the unit cell can be varied by using
different anions (see Figure~\ref{beta}).
Figure~\ref{unitcell} shows the relationship between superconducting
critical temperature $T_{\rm c}$ and hydrostatic pressure
for three $\beta$-phase salts.
It can be seen that increasing the pressure (which
decreases the unit cell size) lowers $T_{\rm c}$;
substitution of a larger anion increases the unit cell
size, countering the effect.

The use of different anions to vary the unit cell size
is often referred to as ``chemical pressure''.
We shall return to the variation of $T_{\rm c}$ with pressure
in Sections~\ref{s2p5} and \ref{s3p1p1}.

\subsection{Intralyer Fermi surface topologies}
The defining property of a metal is that it possesses
a {\it Fermi surface}, that is, a constant-energy
surface in
$k$-space which separates the filled electron
states from empty electron states at absolute zero ($T=0$).
The energy of a quasiparticle\footnote{The term
{\it quasiparticle} refers generally to a
conduction electron or hole modified
by the effects of electron-electron and electron-phonon
interactions~\cite{ashcroft}. We shall discuss this
in more detail in Section~\ref{s2p4}; until
then, the reader is advised to regard quasiparticle
as a general term for an electron or hole in
a solid (see e.g. Reference~\cite{singlebook},
Chapter 8).}
at the Fermi surface
is known as the {\it Fermi energy}, $E_{\rm F}$.
The shape of the Fermi surface is determined
by the dispersion relationships (energy versus
{\bf k} relationships) $E=E({\bf k})$ of each
partially-filled band and the number of
electrons (or holes) to be accommodated.
Virtually all of the properties of
a metal are determined by the quasiparticles
at the Fermi surface, as they occupy states
which are adjacent in energy to empty states;
therefore they are able to respond to
external forces and other perturbations (see e.g.
Chapters 1,2 and 8 of Reference~\cite{singlebook}).

As has been mentioned in the preceding section,
the organic superconductors discussed in this review
are highly anisotropic; most of the quasiparticle
motion occurs within the highly-conducting
planes. The Fermi-surface topology is dominated
by this consideration; the interlayer motion is far
less important. In this section we shall therefore examine the
cross-sections of the Fermi surfaces parallel the highly-conducting planes;
the slight variations in Fermi-surface
cross-sectional area will be discussed in Section~\ref{s2p3}.
\subsubsection{General description}
\label{s2p2p1}
\begin{figure}[htbp]
\centering
\includegraphics[height=6cm]{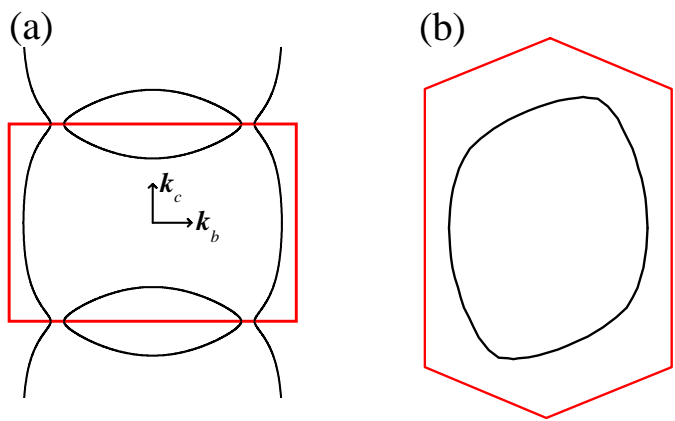}
\caption{(a)~Brillouin zone and Fermi-surface of
$\kappa$-(BEDT-TTF)$_2$Cu(NCS)$_2$,
showing the open, quasi-one-dimensional sections,
and the closed, quasi-two-dimensional
pocket~\cite{schmalian,caulfield,goddard}. (b)~Brillouin zone and Fermi-surface of
$\beta$-(BEDT-TTF)$_2$IBr$_2$~\cite{kartsovnik}.
}
\label{fspic}
\end{figure}

Figure~\ref{fspic}(a) shows a section
(parallel to the highly-conducting planes) through the
first Brillouin zone and Fermi surface of
$\kappa$-(BEDT-TTF)$_2$Cu(NCS)$_2$~\cite{schmalian,caulfield,goddard}.
The Fermi surface may be understood by examining
Figure~\ref{kappapic}(b)~\cite{copout}.
Note that the BEDT-TTF molecules are packed in
pairs called {\it dimers}; if we consider each dimer
as a unit, it will be seen that it is surrounded
by four other dimers.
We thus expect the bandstructure to be fairly isotropic
in the plane,
because there will be substantial,
similar transfer integrals
in these four directions.
The unit cell contains two dimers, each of which
contributes a hole (in the charge-transfer process,
two BEDT-TTF molecules jointly donate one electron to the anion).
Thus, to a first approximation, the Fermi surface for holes might be
expected to be roughly circular with the
same area as the Brillouin zone (as there are two holes
per unit cell)\footnote{The Brillouin-zone is
the primitive unit cell of the reciprocal
($k$-space) lattice. The $k$-space size of the Fermi surface
follows from the properties of phase-space.
Each quasiparticle donated by a unit cell contributes
a volume to the Fermi surface equivalent to
half the volume of the Brillouin zone.
See e.g. Reference~\cite{singlebook},
Chapters 1, 2, 8 and Appendix A,
or Reference~\cite{ashcroft}.}.
As can be seen from Figure~\ref{fspic}(a), the Fermi surface
is {\it roughly} like this; however, the Fermi surface intersects
the Brillouin zone boundaries in the {\bf c} direction,
so that band gaps open up (see e.g. Chapter 2 of
Reference~\cite{singlebook}).
The Fermi surface thus splits into open (electron-like)
sections (often known as {\it Fermi sheets}) running down two of the
Brillouin-zone edges and a closed hole pocket (referred to as
the ``$\alpha$ pocket'')
straddling the other;
it is customary to label such sections
``quasi-one-dimensional'' and
``quasi-two-dimensional'' respectively.
The names arise because the group velocity
${\bf v}$ of the electrons is given
by~\cite{singlebook,ashcroft}\footnote{This
equation is the three-dimensional analogue of the
well-known formula for the group velocity of
a wave packet propagating in one dimension,
$v=({\rm d}\omega /{\rm d} k)$.}
\begin{equation}
\hbar {\bf v} = \nabla_{\bf k} E({\bf k}).
\label{velocities}
\end{equation}
The Fermi surface is a surface of constant
energy; Equation~\ref{velocities}
shows that the velocities of electrons at the
Fermi surface will be directed perpendicular to it.
Therefore, referring to Figure~\ref{fspic},
electrons on the closed Fermi-surface pocket
can possess velocities which point in
any direction in the ($k_b,k_c$) plane;
they have freedom of movement
in two dimensions and are said to be
{\it quasi-two-dimensional}.
By contrast, electrons on the open sections
have velocities predominently directed parallel
to $k_b$ and are {\it quasi-one-dimensional}.

$\kappa$-phase BEDT-TTF superconductors $\kappa$-(BEDT-TTF)$_2X$
can be made with a variety
of other anion molecules, including
$X=$ Cu[N(CN)$_2$]Br (11.8~K), Cu[N(CN)$_2$]Cl (12.8~K (under pressure)),
and I$_3$ (4~K); here the number in parentheses represents $T_{\rm c}$.
In all of these salts, the Fermi-surface topology is very similar
to that in Figure~\ref{fspic}(a); small differences in the symmetry
of the anion layer lead to variations in the gap between the quasi-one-dimensional and quasi-two-dimensional
Fermi-surface sections~\cite{review,mori,ishiguro}.
A summary of the detailed differences and effective
masses\footnote{The effective mass parameterises the
way in which a quasiparticle can respond to an
external force. As we shall see below,
the effective mass also gives a measure
of the density of quasiparticle states at
the Fermi energy. See e.g. Reference~\cite{ashcroft}
or Reference~\cite{singlebook}, Chapter~5.}
is given in Section 3.2 of
Reference~\cite{review} (see also \cite{mori}).

Figure~\ref{fspic}(b) shows the Fermi-surface topology and
Brillouin zone of $\beta$-(BEDT-TTF)$_2$IBr$_2$~\cite{kartsovnik}.
In this case (see Figure~\ref{beta}) there is one hole
per unit cell, so that the Fermi surface cross-sectional area
is half that of the Brillouin zone; only a
quasi-two-dimensional pocket is present.

We have therefore seen that the bandstructure
of a charge-transfer salt is chiefly determined by the
packing arrangement of the cation molecules.
As sources of superconductors, the
$\beta$, $\kappa$, $\beta ''$, $\lambda$ and $\alpha$
phases have been the most important.
The latter four phases all have predicted Fermi surfaces
consisting of a quasi-two-dimensional pocket plus a pair of quasi-one-dimensional Fermi sheets
(the pocket arrangement differs from phase to phase)~\cite{review,wosnitza,mori};
the $\beta$-phase is alone in possessing a Fermi surface
consisting of a single quasi-two-dimensional pocket~\cite{ishiguro}.
\subsubsection{Bandstructure calculations}
The bandstructures of organic superconductors have usually
been calculated using the extended H\"{u}ckel (tight-binding)\footnote{Simple
introductions to the tight-binding model of bandstructure
are given in References~\cite{singlebook,ashcroft}.} approach,
which employs the highest occupied molecular orbitals (HOMOs) of the
cation molecule~\cite{mori}. Section~5.1.3 of Reference~\cite{ishiguro}
discusses this approach and cites some of the most relevant papers.
Whilst this method is usually quite successful in predicting the main
features of the Fermi surface ({\it e.g.} the fact that
there are quasi-one-dimensional and quasi-two-dimensional Fermi-surface sections), the details
of the Fermi-surface topology are sometimes
inadequately described (see {\it e.g.}~\cite{eva}).
This can be important when, for example, the detailed corrugations
of a Fermi sheet govern the interactions
which determine its low-temperature groundstate~\cite{schmalian,eva}.
A possible way around this difficulty is to make slight adjustments of the
transfer integrals so that the predicted Fermi surface is in good
agreement with experimental measurements~\cite{schmalian,caulfield,goddard,eva}.
In the $\beta''$ and $\lambda$ phases the predicted bandstructure
seems very sensitive to the choice of basis set, and
the disagreement between calculation and measurement is often most severe
(see {\it e.g.}~\cite{betawos,doporto1,house,mielke}).

More sophisticated Hubbard-unrestricted Hartree-Fock band calculations
have been carried out for \cuscn ~\cite{demiralp}.
These calculations attempt to take into account many-body effects,
and are successful in reproducing a number of experimental
properties. They also indicate the importance of both
antiferromagnetic fluctuations and electron-phonon interactions
in \cuscn , a fact to which we shall return in
Sections~\ref{s3p1p1} and \ref{s3p4}.
\subsubsection{Experimental measurements of Fermi-surface topology.}
The low scattering rates in quasi-two-dimensional organic superconductors
mean that it is possible to use a range of techniques
to make accurate measurements of the
Fermi-surface topology. The experimental techniques
have been described in other reviews~\cite{review,wosnitza,ishiguro,schrama,amro,amro2}.
We shall mention a few of the more important examples.

\noindent
{\bf de Haas-van Alphen and Shubnikov-de Haas oscillations.}
In a magnetic field, the motion of quasiparticles becomes
partially quantised according to the
equation~\cite{singlebook,ashcroft}
\begin{equation}
E({\bf B},k_z,l)=\frac{\hbar e |{\bf B}|}{m^*}(l+\frac{1}{2})
+E(k_z).
\label{landau}
\end{equation}
Here $E(k_z)$ is the energy of the (unmodified) motion
parallel to {\bf B}, $l$ is a quantum number ($0,1,2,\dots$)
and $m^*$ is an orbitally-averaged effective mass.
The magnetic field quantises the motion of the
quasiparticles in the plane perpendicular to {\bf B};
the resulting levels are known as {\it Landau levels},
and the phenomenon is called {\it Landau quantisation}.
The Landau-level energy separation
is given by $\hbar$ multiplied by
the angular frequency $\omega_c=\hbar eB/m^*$;
this is known as the {\it cyclotron frequency}
because it corresponds to the semiclassical
frequency at which the quasiparticles orbit
the Fermi surface~\cite{singlebook,ashcroft}.

\setcounter{footnote}{1}
{\it Magnetic quantum
oscillations}~\cite{review,wosnitza,shoenberg}\footnote{Introductions
to magnetic quantum oscillations, including the de Haas-van Alphen
and Shubnikov-de Haas effects are given in Chapter 8 of
Reference~\cite{singlebook}, or in Reference~\cite{ashcroft}.}
are caused by the Landau levels passing through the Fermi
energy\footnote{Strictly it is the {\it chemical potential}
$\mu$, rather than the Fermi energy, which is important here.
However $\mu \equiv E_{\rm F}$ at $T=0$, and
$\mu \approx E_{\rm F}$ for virtually all experimental
temperatures of interest~\cite{singlebook,ashcroft}.}.
This results in an oscillation of the electronic
properties of the system, periodic in $1/|{\bf B}|$.
From an experimental standpoint, the oscillations are usually
measured in the magnetisation (de Haas-van Alphen effect)
or the resistivity (Shubnikov-de Haas effect)~\cite{review,wosnitza,shoenberg}.

\begin{figure}[htbp]
\centering
\includegraphics[height=6cm]{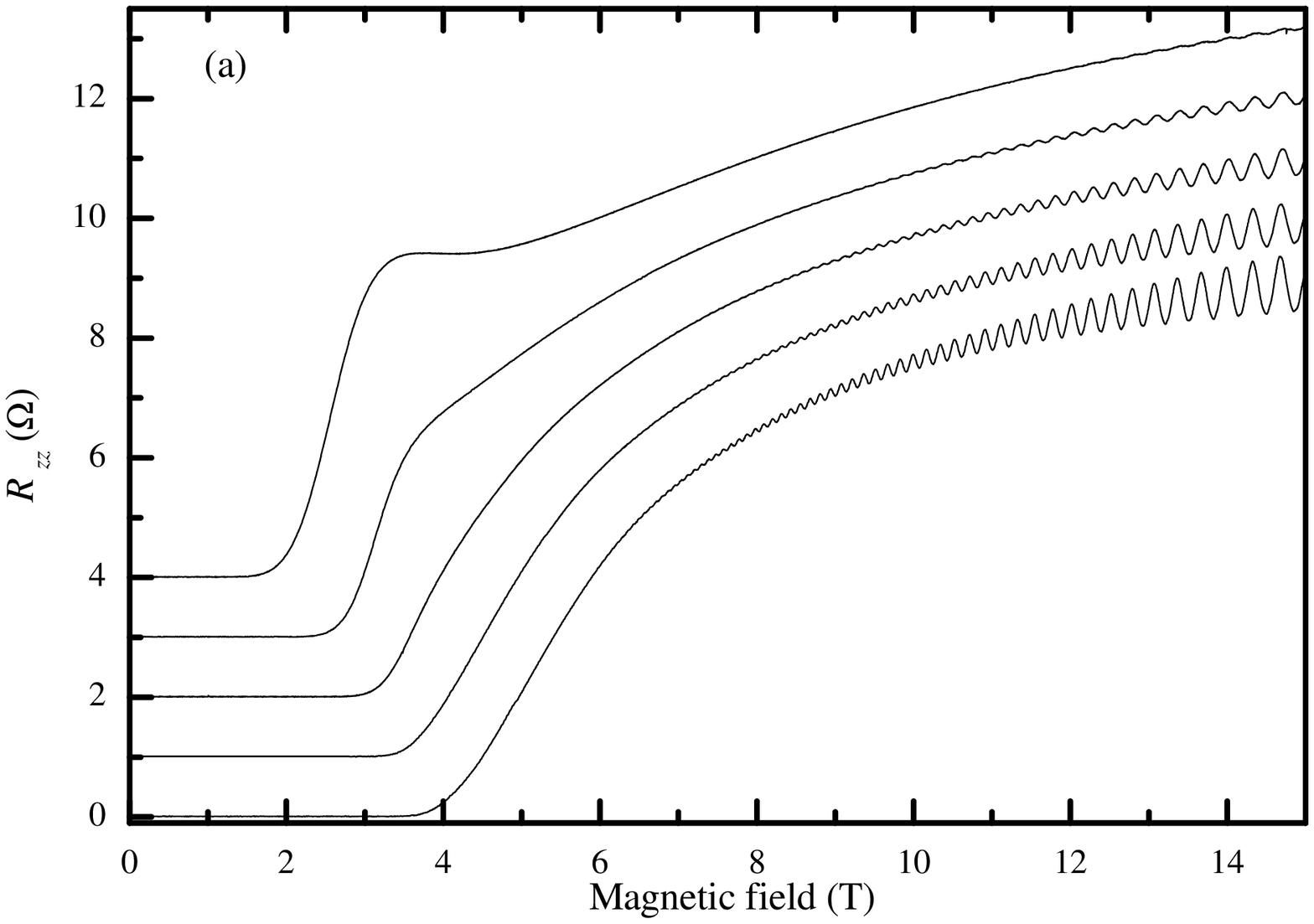}
\includegraphics[height=6cm]{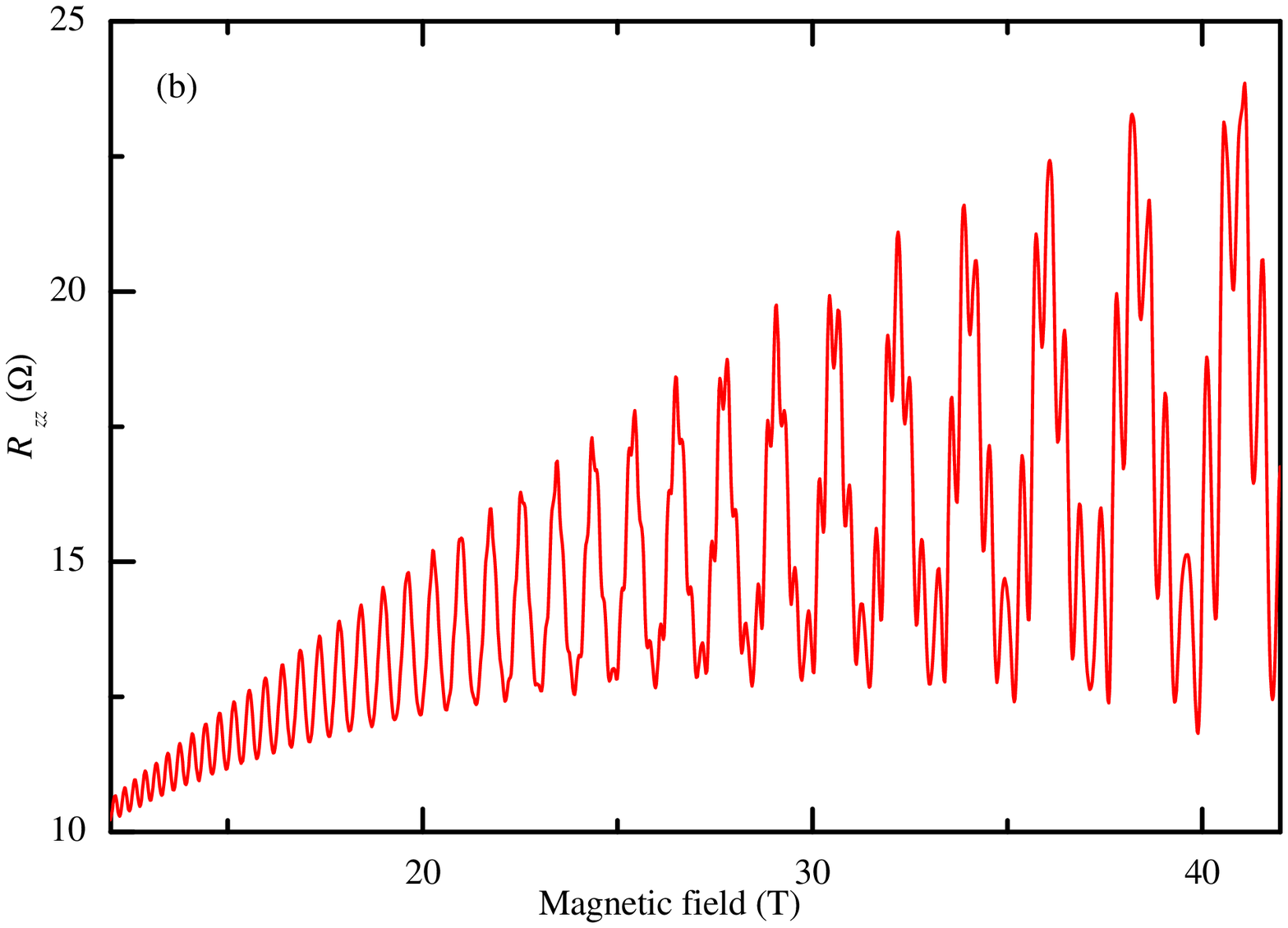}
\caption{Magnetoresistance measurements of
the organic superconductor $\kappa$-(BEDT-TTF)$_2$Cu(NCS)$_2$
(magnetic field applied perpendicular to the quasi-two-dimensional planes).
(a)~Low-field measurements, showing the superconducting to
normal transition and, at higher fields, Shubnikov-de Haas
oscillations caused by the quasi-two-dimensional $\alpha$ pocket of the
Fermi surface. Data for temperatures 1.96~K (uppermost trace),
1.34~K, 1.03~K, 800~mK and 620~mK (lowest trace)
are shown; for clarity, the data have been offset by
$1~\Omega$~\cite{janeisotope}.
(b)~High-field experiment, showing higher frequency
oscillations due to magnetic breakdown ($T=480$~mK)~\cite{hybridpuff}.
In both cases, the current is applied in the
interlayer direction, so that the measured $R_{zz}$
is proportional to the interlayer resistivity component
$\rho_{zz}$~\cite{review}.
Similar data can be found
in {\it e.g.} References~\cite{review,earlycuncs,neilbd}.
}
\label{sdh}
\end{figure}

Landau quantisation only occurs for sections of Fermi surface
corresponding to semiclassical {\it closed} $k$-space
orbits in the plane perpendicular to {\bf B};
the frequency of the oscillation (in Tesla) is given by
$F=(\hbar/2 \pi e)A$, where $A$ is the cross-sectional
$k$-space area of the orbit~\cite{review,wosnitza,shoenberg}.
An example is given in
Figure~\ref{sdh}, which shows Shubnikov-de Haas oscillations in
the magnetoresistance of \cuscn ~\cite{janeisotope,hybridpuff}
(the field was applied perpendicular to the quasi-two-dimensional planes).
Turning first to Figure~\ref{sdh}(a),
a single frequency of oscillations is observed,
caused by the quasi-two-dimensional $\alpha$ pocket of the Fermi surface.
The frequency ($600 \pm 3$~T in this case)
allows one to deduce the cross-sectional
area of the $\alpha$ pocket~\cite{review,wosnitza}.
On increasing the temperature, the oscillations decrease in amplitude,
owing to the thermal smearing of the
Fermi-Dirac distribution function~\cite{review,singlebook}.
The temperature dependence of the oscillation amplitude
can be used to derive the orbitally-averaged effective mass
of the quasiparticles orbiting the $\alpha$ pocket~\cite{review,wosnitza} ($3.5 \pm 0.1 m_{\rm e}$
in this case~\cite{caulfield,janeisotope,earlycuncs}).
Finally, the oscillations increase in amplitude with increasing
magnetic field; this field dependence allows one to derive the
scattering time $\tau$~\cite{review} (3~ps for the data shown~\cite{janeisotope}).

At higher magnetic fields (Figure~\ref{sdh}(b)), an
additional set of higher frequency oscillations becomes
observable. These are due to a phenomenon
known as {\it magnetic breakdown};
the cyclotron energy of the quasiparticles becomes high enough
for them to tunnel (in $k$-space) across the gaps
between the quasi-two-dimensional pocket and the quasi-one-dimensional sheets, so that a
semiclassical orbit around the whole Fermi surface
(a so-called $\beta$ orbit) can be completed~\cite{review,neilbd}.
The large cross-sectional area of this orbit
gives rise to a higher Shubnikov-de Haas oscillation frequency.

\noindent
{\bf Cyclotron resonance.}
Photons of frequency $\nu$ can be
used to excite quasiparticles between Landau levels
(see Equation~\ref{landau})~\cite{review,singlebook};
the phenomenon is known as {\it cyclotron resonance}.
The fundamental resonance occurs when
$\hbar \omega_{\rm c}=h \nu$;
for laboratory fields ($B \sim 1-20$~T)
and the typical effective masses observed in organic
superconductors~\cite{review},
this corresponds to photon frequencies in
the millimetre-wave range, $\nu \sim 10-100$~GHz.
We shall discuss cyclotron resonance
in more detail in Section~\ref{s2p4}.

\noindent
{\bf Angle-dependent magnetoresistance oscillations and Fermi-surface-traversal resonances.}
Whilst they give very accurate information about the cross-sectional
areas of the Fermi-surface sections, magnetic quantum oscillations
do not provide any details of their {\it shape}.
Such information is usually derived from angle-dependent magnetoresistance
oscillations (AMROs)~\cite{review,wosnitza,amro,amro2}.
AMROs are measured by rotating a sample in a fixed magnetic field
whilst monitoring its resistance; the coordinate used to denote
the position of AMROs is the polar angle $\theta$ between the
normal to the sample's quasi-two-dimensional planes and the magnetic field~\cite{amro,amro2}.
It is also very informative to vary the plane of rotation of
the sample in the field; this is described by the azimuthal angle $\phi$~\cite{amro,amro2}.

\begin{figure}[htbp]
\centering
\includegraphics[height=12cm]{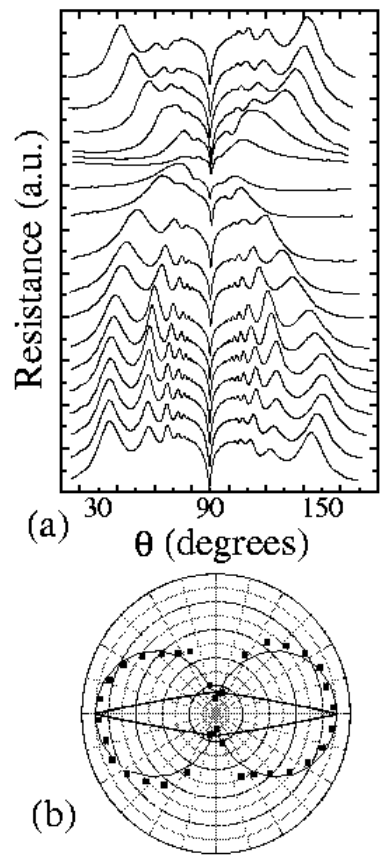}
\caption{(a) AMRO data for
$\beta''$-(BEDT-TTF)$_2$SF$_5$CH$_2$CF$_2$SO$_3$ at 10~T and
1.5~K for $\phi$-angles $7\pm 1^{\circ}$ (top trace)
$17\pm 1^{\circ}$, $27 \pm 1^{\circ}$,
.....$177\pm 1^{\circ}$ (bottom trace-
adjacent traces spaced by
$10\pm 1^{\circ}$).
$\phi=0$ corresponds to
rotation in the {\bf a* c*} plane of the crystal
to within the accuracy of the infrared orientation used.
(b)~The $\phi$ dependence of the
$\tan \theta$
periodicity of the AMRO in (a)
(points); the ``figure of eight'' solid
curve is a fit. The resulting fitted FS
pocket (elongated diamond shape) is shown
within. The long axis of the pocket makes an angle
of $68 \pm 4^{\circ}$ with the ${\bf b}^{\ast}$ axis.
(After Reference~\cite{amro}.)
}
\label{amros}
\end{figure}

AMROs result from the averaging effect that the semiclassical orbits
on the Fermi surface have on the quasiparticle velocity.
Both quasi-one-dimensional and quasi-two-dimensional Fermi-surface sections can give rise to AMROs;
in the former case, the AMROs are sharp dips in the resistivity,
periodic in $\tan \theta$;
in the latter case, one expects {\it peaks}, also periodic in $\tan \theta$~\cite{amro,amro2}.
In order to distinguish between these two cases, it is necessary to carry out
the experiment at several different $\phi$.
The $\phi$-dependence of the AMROs can be related directly to the
shape of a quasi-two-dimensional Fermi-surface section; in the case of a quasi-one-dimensional sheet,
the AMROs yield precise information about the sheet's
orientation~\cite{review,wosnitza,amro,amro2}.
Typical data are shown in Figure~\ref{amros}.

Recently, a high-frequency variant of AMROs, known as the {\it Fermi-surface-traversal
resonance} (FTR), has been developed. This technique allows additional information about
the topology and corrugations of quasi-one-dimensional Fermi sheets to be deduced
(see Reference~\cite{schrama} for a review).

\subsection{\bf 2D or not 2D? Measurements of the effective Fermi-surface dimensionality}
\label{s2p3}
We remarked in Sections~\ref{s2p1} and \ref{s2p2p1}
that the electronic properties
of quasi-two-dimensional organic superconductors are very anisotropic.
The band-structure-measuring techniques mentioned thus far chiefly
give information about the intralayer topology of the Fermi surface.
However, it is important to ask whether the Fermi surface is
exactly two-dimensional, or whether it extends in the interlayer direction,
i.e. is three-dimensional.

This question is of quite general interest, as
there are many correlated-electron
systems which
have very anisotropic electronic bandstructure.
In addition to the organic superconductors~\cite{review,strong},
examples include the ``high-$T_{\rm c}$'' cuprates~\cite{cuprates,strong},
and layered ruthenates~\cite{ruthenate} and manganites~\cite{ramirez}.
Such systems may be described by a tight-binding Hamiltonian
in which the ratio of the interlayer transfer integral $t_{\perp}$
to the average intralayer transfer integral $t_{||}$ is $\ll
1$~\cite{review,strong,mck}.
The inequality $\hbar/ \tau > t_{\perp}$~\cite{mott}
where $\tau^{-1}$ is the quasiparticle
scattering rate~\cite{cuprates,strong,mck},
frequently applies to such systems, suggesting that
the quasiparticles scatter more frequently than they tunnel between layers.
Similarly, under standard laboratory conditions, the inequality
$k_{\rm B}T > t_{\perp}$ often holds, hinting that
thermal smearing will wipe out details of the interlayer periodicity~\cite{anderson}.

The question has thus arisen as to whether the interlayer charge
transfer is coherent or incoherent in these materials,
i.e. whether or not the Fermi surface extends in the interlayer
direction~\cite{review,strong,mck}.
Incoherent interlayer transport is used as a justification
for a number of theories which are thought
to be pivotal in the understanding of
reduced-dimensionality materials (see e.g. \cite{strong,anderson}).
Moreover, models for unconventional superconductivity in $\kappa$-phase
BEDT-TTF salts invoke the nesting\footnote{{\it Nesting}
refers to a topological similarity between
two sections of Fermi surface which allows
one to be translated (by a so-called
{\it nesting vector}) so that it exactly
overlaps with the other. See
e.g. References~\cite{ishiguro,eva}.}
properties of the
Fermi surface~\cite{aoki,schmalian,charffi};
the degree of nesting might depend on whether
the Fermi surface is a two dimensional or three dimensional
entity (see \cite{review},
Section 3.5).

Many apparently solid experimental tests for coherence
in organic superconductors have been deemed to be inconclusive~\cite{mck};
e.g. semiclassical
models can reproduce AMRO~\cite{amro} and FTR data~\cite{schrama}
equally well when the interlayer transport
is coherent or ``weakly coherent''~\cite{mck}.

\begin{figure}
\vspace{80mm}
\includegraphics{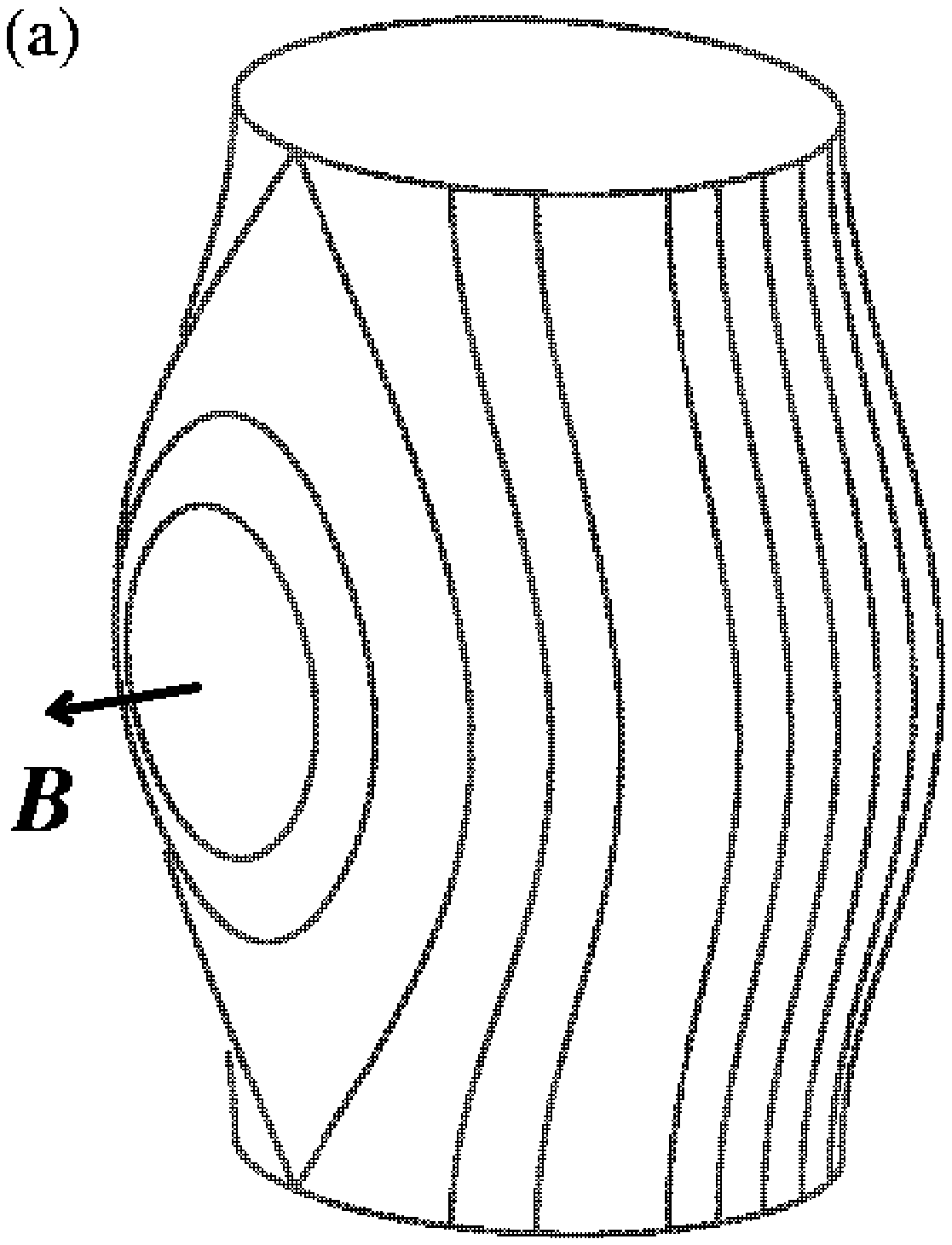}
\includegraphics{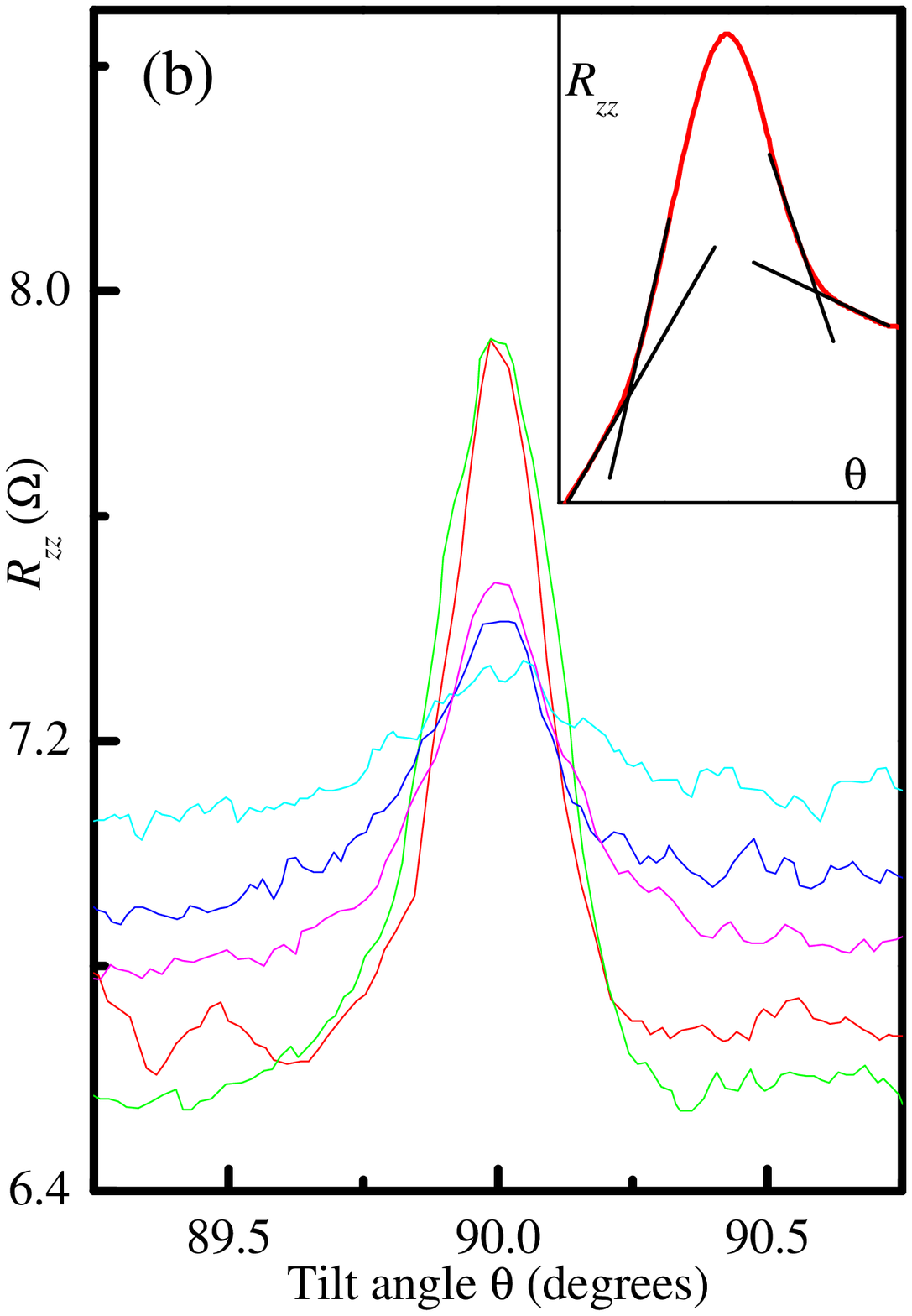}
\caption{(a)~Illustration of the effect of introducing interlayer
dispersion on a quasi-two-dimensional Fermi-surface section; the cross-section of
the Fermi surface varies in the interlayer ($k_{z}$) direction.
The lines on the sides of the Fermi surface illustrate some
of the quasiparticle orbits caused by an in-plane field.
(b)~Interlayer resistance $R_{zz}$ (proportional to $\rho_{zz}$)
of a \cuscn ~sample as a function of magnetic-field
orientation ($\theta=90^{\circ}$ corresponds to magnetic field
exactly in-plane). Data for temperatures
$T = 0.48$~K, 1.4~K, 3.0~K, 4.4~K and 5.1~K
are shown.
The background magnetoresistance increases
with increasing $T$, whereas the peak at $\theta =90^{\circ}$
becomes smaller.
The inset shows the intersections of the linear extrapolations
used to determine the peak width (after Reference~\cite{goddard}.
}
\label{belly}
\end{figure}

A simple tight-binding
expression is often used to simulate the interlayer ($z$-direction)
dispersion~\cite{review};
$E(k_z)=-2t_{\perp}\cos (k_za)$.
Here $t_{\perp}$ is the interlayer
transfer integral and $a$ is the unit-cell height in the $z$ direction.
The introduction of the interlayer dispersion causes
a modulation of the Fermi-surface cross-section,
shown schematically in Figure~\ref{belly}(a)~\cite{footnote,cokebottle}.
If the Fermi surface is extended in the $z$ direction, a magnetic field
applied exactly in the intralayer ($xy$) plane
can cause a variety of orbits on the sides of the Fermi surface,
as shown schematically in Fig.\ref{belly}(a).\footnote{The
orbits are defined by the Lorentz force
$\hbar({\rm d}{\bf k}/{\rm d}t)=-e{\bf v} \times {\bf B}$,
where {\bf v} is given by Equation~1.
This results in orbits on the Fermi surface,
in a plane perpendicular to {\bf B}~\cite{review,ashcroft}.}
It has been proposed that the closed orbits
about the belly of the Fermi surface~\cite{hanasaki}
or the ``self-crossing orbits'' found under
the same conditions~\cite{russian}
are very effective in averaging $v_{\perp}$,
the interlayer component of the velocity.
Therefore, the presence of such orbits
will lead to an increase in the resistivty
component $\rho_{zz}$~\cite{hanasaki,russian}.
On tilting {\bf B} away from the intralayer direction,
the closed and ``self-crossing'' orbits cease to be possible when
{\bf B} has turned through an angle $\Delta$,
where
\begin{equation}
\Delta ({\rm in~radians}) \approx v_{\perp}/v_{||}.
\label{delta}
\end{equation}
Here $v_{\perp}$ is the maximum of the interlayer
component of the quasiparticle velocity, and $v_{||}$ is
the intralayer component of the quasiparticle velocity
in the plane of rotation of {\bf B}.
Therefore, on tilting {\bf B} around the in-plane orientation,
one expects to see a peak in $\rho_{zz}$, of angular width $2 \Delta$,
if (and only if~\cite{mck}) the Fermi surface is extended in the $z$ direction.
By using Equations~\ref{velocities} and \ref{delta}
and measured details of the
intralayer Fermi-surface topology, it is possible to
use $\Delta$ to deduce $t_{\perp}$~\cite{goddard}.

Figure~\ref{belly}(b) shows typical data for \cuscn .
The observation of a peak in
$\rho_{zz}$ close to $\theta =90^{\circ}$ allows the
interlayer transfer integral to be estimated to be
$t_{\perp} \approx 0.04$~meV~\cite{goddard}.
This may be compared with intralayer transfer integrals $\sim 150$~meV~\cite{caulfield}.
Such data are of great interest because they illustrate that the
criteria used to delineate interlayer incoherence are rather a poor
guide to real samples. For example,
the sample shown in Figure~\ref{belly}(b) has
$\hbar/\tau \approx 6 t_{\perp}$, and yet robustly shows
evidence for a three-dimensional Fermi surface~\cite{goddard}.
Similarly, a temperature of $5$~K, ($k_{\rm B}T \approx 10 t_{\perp}$)
leads one to expect
incoherent interlayer transport~\cite{anderson},
yet the peak in $\rho_{zz}$ shown in Figure~\ref{belly}(b) unambiguously
demonstrates a three dimensional Fermi-surface topology.

Such measurements have been carried out on the organic
superconductors $\beta$-(BEDT-TTF)$_2$IBr$_2$~\cite{kartsovnik},
$\kappa$-(BEDT-TTF)$_2$Cu$_2$(CN)$_3$~\cite{ohmichi} (under pressure),
\nh4 ~\cite{hanasaki} (under pressure), $\beta$-(BEDT-TTF)$_2$I$_3$~\cite{hanasaki},
\cuscn ~\cite{goddard},
\lbets ~\cite{mielke} and \betasf ~\cite{wosnisquit,janesquit}.
In the latter example, no peak was observed, suggesting incoherent
interlayer transport.
In all of the other instances, the $\rho_{zz}$ data demonstrate a
Fermi surface which is extended in the interlayer direction.
\subsection{Renormalising interactions.}
\label{s2p4}
A simple bandstructure calculation essentially
uses ions and molecules
which are rigidly fixed in a perfectly
periodic arrangement to obtain a periodic
potential and hence the bands.
However, the ions and/or molecules in
a substance will in general be charged,
or at the very least possess a dipole moment;
as an electron passes through the solid,
it will tend to distort the lattice around it owing to
the Coulomb interactions between the ions and molecules
and its own charge.
This leads to the electron being accompanied by
a strain field as it moves through the substance;
alternatively one can consider the electron
being surrounded by virtual phonons,
leading to a change in the electron's
self energy~\cite{ashcroft}. Similarly, self-energy effects due to
the interactions between the
conduction electrons themselves must be taken
into account~\cite{ashcroft,singleton,toyota}.

The inclusion of such effects leads to the idea of
``quasiparticles'', excitations of an interacting
electron system which obey Fermi-Dirac statistics
and which ``look like'' conduction electrons or holes in many
ways, but which possess renormalised effective
masses and scattering rates~\cite{ashcroft,toyota}.\footnote{A collection
of interacting electrons which can be treated
(using Fermi-Dirac statistics) as a collection of quasiparticles is
often known as a {\it Fermi liquid}; by contrast, a collection of
non-interacting electrons is called a {\it Fermi gas}~\cite{ashcroft}.
At present, it appears that most of the normal-state
properties of quasi-two-dimensional organic superconductors are fairly well
described by Fermi-liquid theory~\cite{review,wosnitza,ishiguro}.}
In the presence of relatively weak interactions,
it is possible to make predictions about the
contribution to the quasiparticle effective mass caused by
the above effects~\cite{singleton,quader}.
If $m_{\rm b}$ is the
{\it band mass} calculated in the bandstructure
calculations, then the electron-lattice interactions
discussed above will result in a
{\it dynamical mass}, $m_{\lambda}$, where
\begin{equation}
m_{\lambda} \approx (1+\lambda)m_{\rm b},
\end{equation}
where $\lambda$ is an electron-phonon coupling
constant (in BEDT-TTF salts, $\lambda$
is thought to be between 0.1 and 1~\cite{ko}).
Finally, the electron-electron interactions result in
an {\it effective mass}
$m^*$;
\begin{equation}
m^*\approx (1+\frac{F^1_{\rm s}}{3})m_{\lambda},
\end{equation}
where $F^1_{\rm s}$ is a constant known as
a Fermi liquid parameter~\cite{singleton,quader}.

How do such considerations affect experimental data?
The interactions which give rise to the effective mass renormalisation
chiefly affect states very close to the Fermi energy. Therefore,
an interband optical measurement,
which probes the full energy range of
a band, is detemined mostly by its bare (unrenormalised) width,
enabling $m_{\rm b}$ to be extracted~\cite{caulfield}.
On the other hand, a cyclotron resonance experiment
is expected to be relatively insensitive to interactions involving
relative motion of electrons (i.e. the electron-electron interactions).
The reason for this is that the wavelength of the
photons used to excite cyclotron resonance ($\sim 1$~mm)
is much greater than the typical quasiparticle spacing ($\sim 10$~\AA),
so that adjacent electrons experience an oscillatory electric
field of the same phase; their relative
separation is therefore not affected.
Hence, the mass measured in cyclotron resonance will
be close to the (orbitally averaged) dynamical
mass, $m_{\lambda}$~\cite{quader}. Finally, the (orbitally averaged)
effective mass (including the contributions of both electron-electron and
electron-phonon interactions)
is the mass measured in Shubnikov-de Haas and de
Haas-van Alphen experiments~\cite{singleton,quader}.
Following these ideas, Caulfield {\it et al.} compared
optical spectra and Shubnikov-de Haas data in the superconductor
\cuscn ~\cite{caulfield}, finding
$m^*/m_{\rm b} \sim 5$~\cite{caulfield,crdisclaimer} .

\begin{figure}[htbp]
   \centering
   \includegraphics[height=10cm]{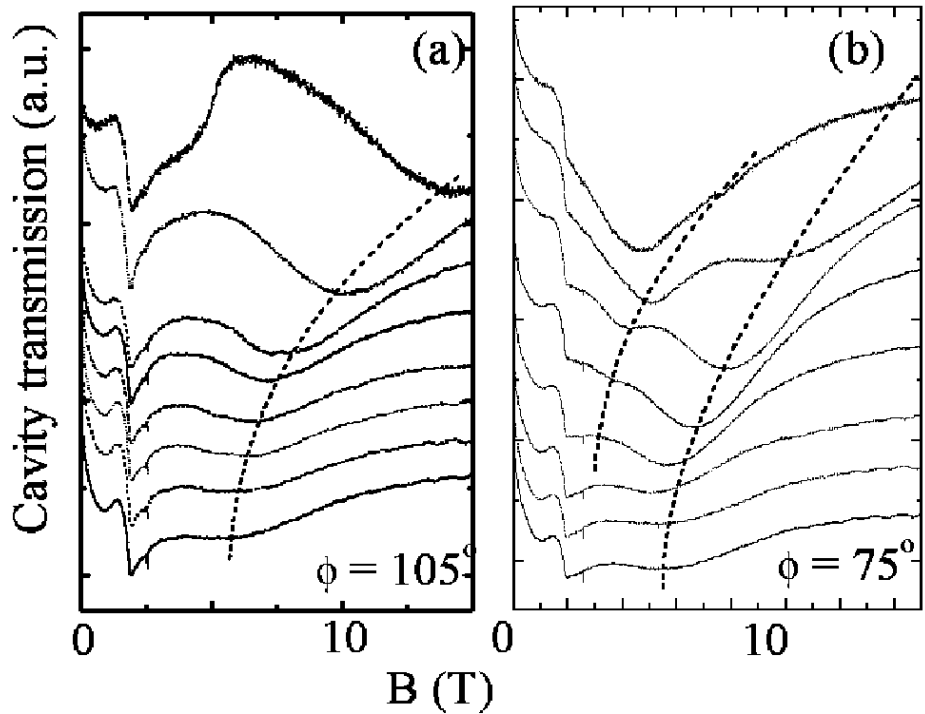}
\caption{Cyclotron resonance in \betasf ~(after Reference~\cite{schrama}).
(a) Transmission of a resonant
cavity loaded with a single crystal of
$\beta^{\prime\prime}$-(BEDT-TTF)$_2$SF$_5$CH$_2$CF$_2$SO$_3$
versus magnetic field for $\theta=0^{\circ}$ (lowest trace) to
$\theta=70^{\circ}$ (uppermost trace) and $\phi
=105^{\circ}$ ($T=1.5$~K).
Here $\theta$ denotes the angle between the
normal to the quasi-two-dimensional planes of the sample and the magnetic field;
$\phi$ refers to the plane of rotation of the sample
within the magnetic field~\cite{schrama}.
(b) Equivalent data for $\phi = 75^{\circ}$.
In both data sets, the structure at low
fields is associated with the superconducting-normal
transition of \betasf , and with a background feature
of the cavity. The dotted lines donate movement of the cyclotron
resonance (a,b) and its second harmonic ((b) only)~\cite{schrama}. The
azimuthal angle dependence of the harmonic intensity
enables the shape of the Fermi surface to be mapped.
}
\label{crpic} 
\end{figure}

These considerations also motivated
a number of attempts to observe
cyclotron resonance in charge-transfer salts~\cite{singleton,etcr}.
For reasons discussed in Section~2.4.6 of Reference~\cite{review},
many of the earlier experiments were seriously flawed, and
reliable experiments had to await the
development of resonant cavity techniques~\cite{ftr,shill,anton}.
Since such measurements became feasible, only
two superconducting salts
have yielded cyclotron resonance data.
In $\alpha$-(BEDT-TTF)$_2$NH$_4$Hg(SCN)$_4$,
which is believed to have a simple Fermi surface
consisting of a pair of sheets and a closed pocket,
a cyclotron resonance corresponding to
a mass of $1.9 m_{\rm e}$
was measured~\cite{anton,shill};
this may be compared with an effective mass
from magnetic quantum oscillations of $2.5 m_{\rm e}$~\cite{marianoprl}.
\betasf,
~again with a Fermi surface believed to consist
of a single pocket and a pair of
sheets (but see Reference~\cite{msnloc}),
exhibited a cyclotron resonance corresponding to
a mass of $2.2m_{\rm e}$~\cite{schrama} (see Figure~\ref{crpic})
the effective mass from Shubnikov-de Haas
oscillations is $1.9m_{\rm e}$~\cite{betabandstructure}.

Thus it appears that the
simple predictions of a large enhancement
of the mass derived from magnetic quantum
oscillations over that measured in a cyclotron resonance
experiment do not hold.
Recently, Kanki and Yamada~\cite{kandy}
have carried out Hubbard-model
calculations which contradict the
simplistic theories~\cite{quader,kohn} and which
appear to support the
above experimental observations.
Kanki and Yamada found that the relationship
between the mass measured in a cyclotron
resonance experiment and that derived from
magnetic quantum oscillations depends
strongly on {\it e.g.} bandfilling,
and in some cases the former can exceed
the latter, as apparently seen in
\betasf~\cite{schrama}.

At any rate, it appears that electron-electron
and electron-phonon interactions are important in
quasi-two-dimensional organic superconductors, leading to
substantial renormalisations of the quasiparticle masses.
\subsection{The pressure dependence of the bandstructure and
its effect on superconductivity.}
\label{s2p5}
The pressure dependence of the magnetic quantum oscillations
has been examined in
$\kappa$-(BEDT-TTF)$_2$Cu(NCS)$_2$~\cite{caulfield,caulfield1},
$\kappa$-(BEDT-TTF)$_2$Cu[N(CN)$_2$]Br~\cite{weiss,weiss1}
(see also comments in Reference~\cite{sdh2}),
$\kappa$-(BEDT-TTF)$_2$Cu[N(CN)$_2$]Cl~\cite{kartsjetp}
and $\kappa$-(BEDT-TTF)$_2$Cu$_2$(CN)$_3$~\cite{ohmichi3}.
The results from all four materials follow the same
general trend, and so we shall illustrate the general features
using data from References~\cite{caulfield,caulfield1}.

\begin{figure}[htbp]
\centering
\includegraphics[height=10cm]{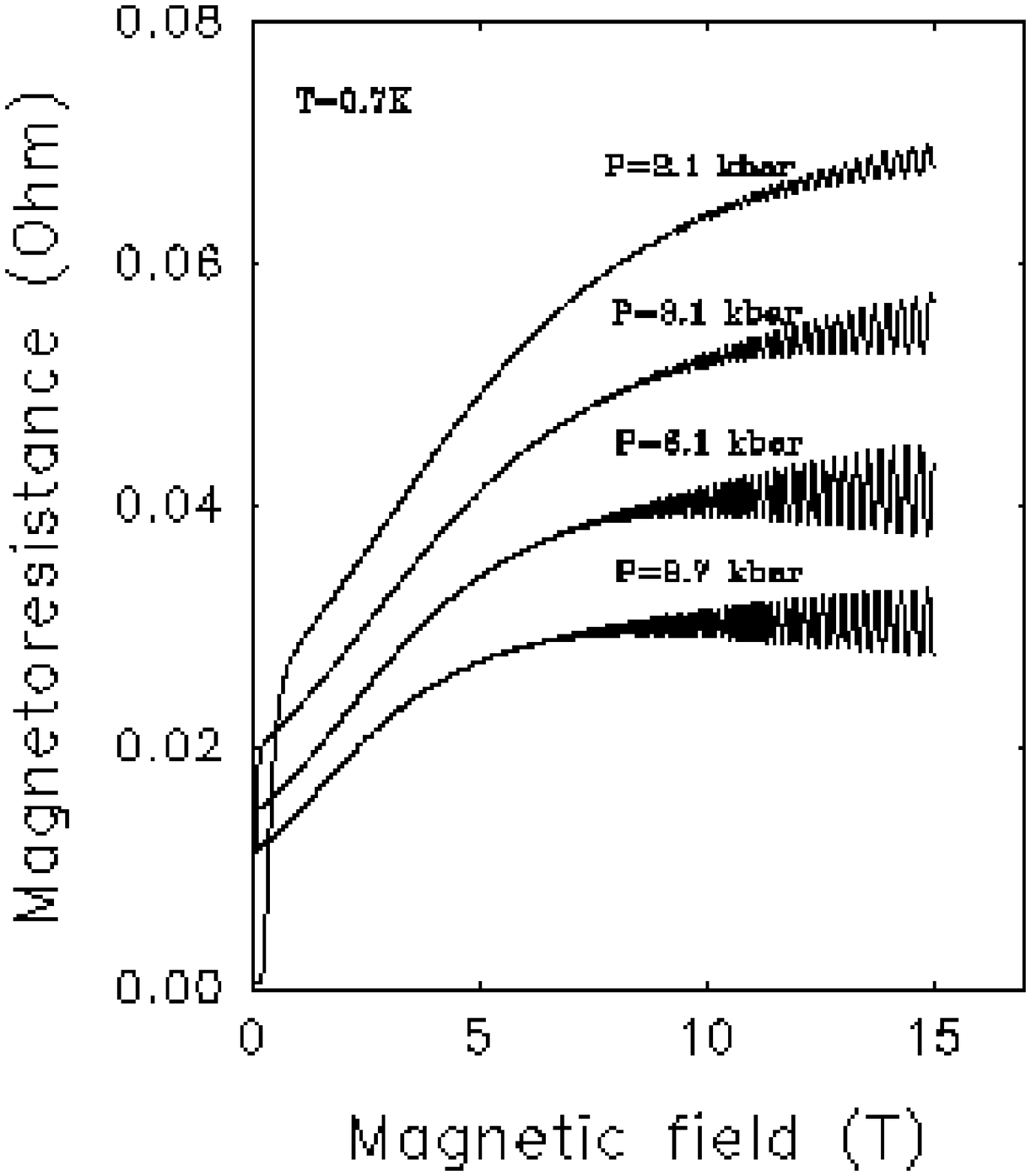}
\caption{Resistance of $\kappa$-(BEDT-TTF)$_2$Cu(NCS)$_2$
at $T=0.7$~K for several different hydrostatic pressures
(after Reference~\cite{caulfield}).}
\label{pressuremr}
\end{figure}

Figure~\ref{pressuremr} shows the
resistance of $\kappa$-(BEDT-TTF)$_2$Cu(NCS)$_2$
at $T=0.7$~K for several different hydrostatic
pressures~\cite{caulfield,caulfield1}.
A comparison with Figure~\ref{sdh} shows that
the increasing pressure suppresses the superconductivity;
by $P=3.1$~kbar, the critical field is a fraction of a Tesla.
At 6.1~kbar there is no evidence of superconductivity
at all.
However, the Shubnikov-de Haas oscillations
due to the $\alpha$ hole pocket
remain present, so that its $k$-space area and
$m^*_{\alpha}$ can be extracted as a function of
pressure.
As the pressure increases, the $\beta$ breakdown oscillations
also become visible at much lower magnetic fields~\cite{caulfield}.

\begin{figure}[htbp]
\centering
\includegraphics[height=10cm]{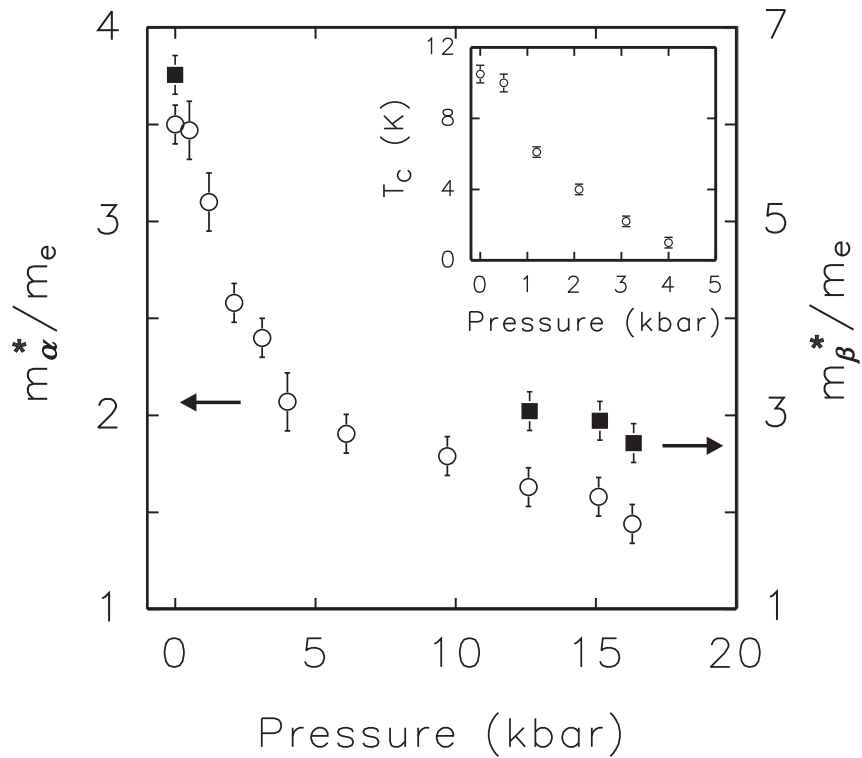}
\caption{Effective masses of the $\alpha$
hole pocket $m^*_{\alpha}$ and $\beta$ orbit $m^*_{\beta}$
as a function of pressure (main Figure).
The inset shows the superconducting critical
temperature $T_{\rm c}$ as a function of pressure
(after Reference~\cite{caulfield}).}
\label{mabtcpt}
\end{figure}

Figure~\ref{mabtcpt} shows the pressure dependence of
$m^*_{\alpha}$ (corresponding to the
$\alpha$ hole pocket), $m^*_{\beta}$ (corresponding
to the $\beta$ orbit))
and the superconducting critical temperature.
The application of pressure causes both masses
to decrease (see also Reference~\cite{weiss1}).
This decrease in effective mass with increasing pressure
seems to be a universal feature of the
$\kappa$-phase BEDT-TTF salts; Weiss {\it et al.}
have shown that the values of $m^*_{\beta}$
for three different $\kappa$-phase salts
map onto a common curve as a function of pressure~\cite{weiss1}.

The superconducting critical temperature decreases
very rapidly as the masses fall;
by 5~kbar, $T_{\rm c}$ is immeasurably small.
It is well known that the density of states at the Fermi
energy is proportional to the effective mass~\cite{caulfield,ashcroft};
as the effective masses fall with increasing pressure,
the density of states at the Fermi energy will
also decrease.
Mechanisms for superconductivity involve the
pairing of electrons of equal and opposite
{\bf k} caused by the
exchange of virtual excitations such as
phonons~\cite{tinkham}.
The strength of the pairing is directly determined by
the rate at which this exchange can take place,
which in turn depends on the density of states
at the Fermi energy~\cite{caulfield1}.
Thus, the decrease of
the effective masses reduces the density of states at the Fermi energy,
thereby suppressing the superconductivity.
Any model of superconductivity should therefore potentially
allow one to relate $T_{\rm c}$ to $m^*$.
Lee solved the linearised Eliashberg equations
to fit the dependence of $T_{\rm c}$ on $m^*$~\cite{caulfield1,lee}
using an Einstein phonon energy of 5~meV
(close to a prominent maximum in the phonon
density of states)
an ambient-pressure electron-phonon coupling constant
of $\lambda \approx 0.4$ ({\it c.f.} values $\lambda \sim 0.3-0.5$
obtained in Reference~\cite{ko})
and an ambient-pressure Coulomb pseudopotential
$\mu^*=-0.22$.
Lee pointed out~\cite{lee} that
``an attractive interaction of high-energy origin
must be responsible for the pairing potential,
which is parameterised by a large negative Coulomb
pseudopotential..... this high-energy
interaction... could come from spin fluctuations.''
Since Lee's work, a number of other authors have
noted the importance of spin fluctuations
in the $\kappa$-phase BEDT-TTF salts (see Section~\ref{s3p1p1}).

\begin{figure}
\centering
\includegraphics[height=8cm]{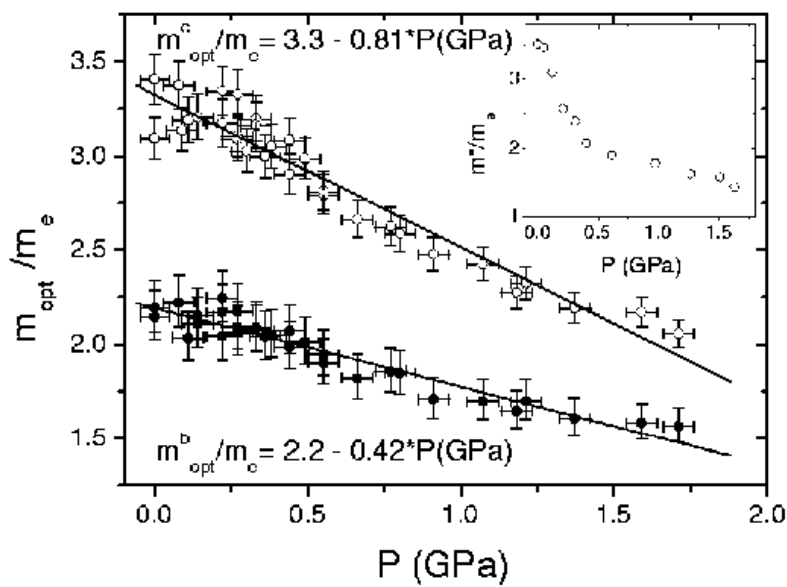}
\caption{``Optical masses'' deduced from reflectivity
data as a function of pressure.
The variation of the effective mass with pressure
from Figure~\ref{mabtcpt} is reproduced in the inset
(after Reference~\cite{klehe}).}
\label{akkmass}
\end{figure}

In order to clarify what is responsible for the
reduction in effective mass with pressure, Klehe {\it et al.}~\cite{klehe}
performed reflectivity measurements on a sample of \cuscn ~as
a function of pressure. They were able to deduce
``optical masses'' (Figure~\ref{akkmass})
which are closely related
to the bare band mass (see previous section).
Although the absolute values of the optical
masses are likely to be scaled by some numerical factor~\cite{klehe},
their relative shift with pressure is a good guide
to what happens to the bare band mass.
In contrast to the ``elbow-like''
variation of the effective mass
with pressure (see Figure~\ref{mabtcpt}),
the optical mass shows a smooth decrease.
This indicates that the rapid change in effective mass
seen below 5~kbar (0.5~GPa) is associated with
a strong decrease in the electron-electron interactions
and/or electron-phonon interactions (which contribute
to the effective mass but not to the bare band mass;
see Section~\ref{s2p4}).

.

``Chemical pressure'', that is, the use of
longer anions to stretch the unit cell, can be used
to explore what are effectively
negative pressures compared to the actual pressures in Figure~\ref{mabtcpt}.
The most ``negative'' chemical pressure is achieved with
$\kappa$-(BEDT-TTF)$_2$Cu[N(CN)$_2$]Cl, which
is an insulator at ambient (actual)
pressure~\cite{kartsjetp,sinojap}. (This salt
is actually what is known as a {\it Mott insulator}~\cite{ishiguro,ashcroft},
in which the transfer integrals are small compared to
Coulomb repulsion between holes on
neighbouring dimers, thus preventing the holes from hopping.)
The chemical pressure is less negative
for $\kappa$-(BEDT-TTF)$_2$Cu[N(CN)$_2$]Br~\cite{weiss,weiss1},
which teeters between insulating and metallic behaviour;
as a result, the detailed behaviour of this salt
has received considerable attention~\cite{tanatar,french}.
We shall return to the effect of pressure and the superconducting
phase diagram below (Figure~\ref{nature}).
\subsection{The temperature dependence of the resistance.}
\label{s2p6}
The normal-state resistivity of the superconducting BEDT-TTF salts
varies as $T^2$, at least up to temperatures $T \sim 30-40$~K~\cite{ishigutsq}.
This is yet another indication that electron-electron interactions
are important~\cite{ashcroft} in quasi-two-dimensional organic superconductors.
Above this temperature, the rate of increase in resistivity
with temperature slackens off.

At slightly higher temperatures still, the resistivities
of the $\kappa$-phase BEDT-TTF salts
tend to exhibit a broad peak or hump.
In the case of
\cuscn , the hump occurs at about 90~K~\cite{ishiguhump}.
Above this temperature, the resistivity of the $\kappa$-phase salts
decreases slowly
with increasing temperature (i.e. they have an ``insulating''-type
behaviour).
The hump is usually associated with small rearrangements
of the terminal ethylene groups of the BEDT-TTF molecule~\cite{ishiguhump};
however, recent optical data have been used to support the argument
that the transition from ``insulating'' (high $T$)
to ``metallic'' (low $T$)
behaviour occurring at the hump represents a shift
from small polaron to large polaron conduction~\cite{wang}.\footnote{The
term {\it polaron} is used to describe an electron accompanied
by a strain field (i.e. a distortion of the crystal caused
by the electron interacting with polar molecules)~\cite{ashcroft}.}

At this point it is necessary to add a note of caution about
resistivity measurements in quasi-two-dimensional organic conductors.
Whilst interlayer resistivity ($\rho_{zz}$)
measurements are simple~\cite{contacts},
the very large resistivity anisotropy $\sim 10^3-10^5$~\cite{review}
makes quantitative experimental measurements
of the intralayer resistivity
using conventional wires and contacts almost impossible~\cite{contacts};
there is almost always a substantial component of $\rho_{zz}$ present.
Part of problem stems from the irregular shape and small
size of the high-quality single crystals;
typical dimensions are $\sim 1 \times 0.5 \times 0.1$~mm$^3$~\cite{contacts};
at present, high quality epitaxial films are not available.

Indeed, it is very probable that the only reliable estimates of the
intralayer resistivity are those derived from
skin-depth (see Reference~\cite{mielke} and references
therein) and optical conductivity studies~\cite{dressel}.
\subsection{Summary of Fermi-surface and quasiparticle properties}
In the following sections, we shall describe what is known
about the superconducting state of quasi-two-dimensional (quasi-two-dimensional) organic
superconductors. It is therefore a convenient point to summarise all that has
been discussed about the normal-state properties thus far.
\begin{enumerate}
\item
The organic superconductors described in this review
are {\it charge-transfer salts}, comprising bandstructure-forming
cation molecules Y (Y is BEDT-TTF, BETS, BEDO etc.)
and anion molecules X; $n$ Y molecules
jointly donate an electron to X, stabilising the charge-transfer salt
Y$_n$X and leaving behind a hole.
\item
The band-structure-forming molecules pack into layers,
separated by layers of anion molecules. The bandshape and bandwidth
can be altered chemically by changing the anion.
The various structural morphologies available are denoted
$\alpha$, $\beta$, $\beta''$, $\kappa$, $\lambda$ {\it etc.}.
\item
The layered nature of the salts results in a quasi-two-dimensional bandstructure.
The extensive application of techniques such as the de Haas-van Alphen effect,
angle-dependent magnetoresistance oscillations and the Fermi-surface traversal
resonance means that the intralayer Fermi-surface topology is
is often known to good accuracy~\cite{review}.
\item
The interlayer transfer integrals of quasi-two-dimensional organic superconductors are
very small ($\sim 0.5-2$~Kelvin). Nevertheless, many appear to
exhibit unambiguous experimental signatures of a three-dimensional
Fermi surface.
\item
Many-body effects are quite pronounced, leading to a Fermi-liquid
picture in which the quasiparticle effective mass
can be several times larger than the bare band mass.
\end{enumerate}

\section{The superconducting state; mechanism for superconductivity}
A discussion of the superconducting state of quasi-two-dimensional organic
conductors is complicated by apparently contradictory
experimental evidence.
Most has been done on the $\kappa$-phase BEDT-TTF
salts, and at first (in the early to mid 1990s) it was
frequently assumed that their superconductivity
could be described by a Bardeen-Cooper-Schrieffer (BCS)-like
model\footnote{The Bardeen-Cooper-Schrieffer (BCS) theory
envisages pairing between quasiparticles of equal and opposite
momentum and spin, with the pairing interaction being
communicated by the exchange of a virtual phonon (lattice vibration).
As a result, an energy gap of size
$2 \Delta$, where $\Delta$ is known as
the {\it order parameter}, opens up around $E_{\rm F}$,
separating the paired states from the normal quasiparticles.
In the BCS model this energy gap is uniform over all the
Fermi surface (this is known as an {\it s-wave-type order
parameter}).
The BCS theory appears to describe elemental superconductors
and many alloys very adequately~\cite{ashcroft}.}
It was assumed that the order parameter $\Delta$
was roughly isotropic
(s-wave-like); at the time,
data from several experiments seemed not
inconsistent with this interpretation
(see {\it e.g.} References~\cite{caulfield,harshman,lang,dresselmm}).
Subsequently, considerable doubt was cast on this idea~\cite{ishikres},
with the greater proportion of more recent
experimental data suggesting
that these salts exhibit a d-wave order parameter (with nodes),
the superconductivity being mediated by antiferromagnetic fluctuations.
In this context, the nodes are regions over which the
order parameter (and energy gap) becomes zero;
d-wave refers to the likeness of the node and antinode arrangement
to atomic d orbitals.

However, the situation is not quite resolved,
as a number of other experiments suggest a completely gapped
superconducting state~\cite{elsinger,Faulques}.
Moreover, infrared reflectivity and Raman data present a
plethora of results and interpretation suggesting that
electron-phonon interactions,
antiferromagnetic fluctuations {\it and}
perhaps other electron-electron interactions are all important
(or at least in some way involved) in the mechanism for
superconductivity, although the details remain unclear,
and in some cases the interpretations are divergent. 

Most of the more recent experiments have been carried out on
the $\kappa$-phase BEDT-TTF salts.
We shall treat these results in some detail, giving a summary
of the behaviour of the $\beta$-phase salts in Section~\ref{s3p4p2}.
\subsection{The $\kappa$-phase BEDT-TTF salts; evidence in favour of d-wave and gap nodes.}
\subsubsection{Proximity of superconductivity to antiferromagnetism; NMR experiments.}
\label{s3p1p1}
\begin{figure}
\centering
\includegraphics[height=8cm]{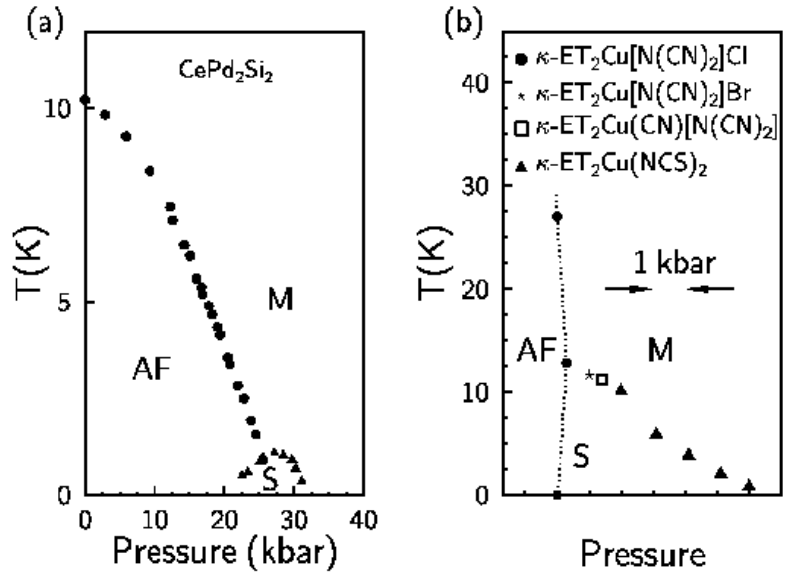}
\caption{An illustration of the similarity between $\kappa$-phase BEDT-TTF
superconductors and heavy-fermion compounds.
(a) Temperature-pressure phase diagram of the
heavy fermion superconductor CePd$_2$Si$_2$
(after Reference~\cite{mathur}).  (AF=antiferromagnetism,
S=superconducitivity, M=metal)
(b) Temperature-pressure phase diagram of the
organic superconductors
$\kappa$-(BEDT-TTF)$_2$Cu(NCS)$_2$,
$\kappa$-(BEDT-TTF)$_2$Cu[N(CN)$_2$]Br,
$\kappa$-(BEDT-TTF)$_2$Cu(CN)[N(CN)$_2$] and
$\kappa$-(BEDT-TTF)$_2$Cu[N(CN)$_2$]Cl.
The pressure axis includes the effect
of ``chemical pressure'' caused by chemically
varying the unit cell size (see Section~\ref{s2p5})
as well as conventionally applied hydrostatic pressure;
the lines indicate the ambient-pressure
positions of the three named substances
(adapted from References~\cite{caulfield,ross,french,kanoda}).
Note that the left-hand region of the phase diagram is
somewhat simplified for clarity~\cite{anguish};
see Figure~\ref{lefphase}~\cite{french}
(After Reference~\cite{review}.)
}
\label{nature}
\end{figure}

Initial proposals for non-phonon-mediated superconductivity
in the $\kappa$-phase BEDT-TTF salts were prompted by similarity between their
phase diagrams and those of
the high-$T_{\rm c}$ cuprates and heavy-fermion compounds~\cite{review,mck1,ross}
(see Figure~\ref{nature}).
It is important
that in all cases, the superconducting region
of the phase diagram is in close proximity to
antiferromagnetism, suggesting that antiferromagnetic
fluctuations are important in the superconducting mechanism.

Such proposals gain some of their strongest support
from nuclear magnetic resonance (NMR)
measurements.
Relaxation rates ($1/T_1$)
and Knight shifts carried out by three
independent groups~\cite{desoto,mayaffre,kawamoto}
show behaviour reminiscent of the
high-$T_{\rm c}$ cuprates, in which antiferromagnetic
phase fluctuations and spin-gap behaviour dominate;
the Hebel-Slichter peak, a feature expected for fully
gapped (BCS-like) superconductors, is absent~\cite{desoto,mayaffre,kawamoto}
(see Reference~\cite{ishikres} and References therein for a general comparison of
the NMR properties of the $\kappa$-phase BEDT-TTF salts with those of other materials).
In the superconducting state,
it has been stated that the NMR data ``rule out the BCS electron-phonon
mechanism as the source of the superconductivity,
but support an unconventional pairing state with possible nodes
in the gap function''~\cite{desoto}; similar comments
are made in References~\cite{mayaffre,kawamoto}, based on the
observation that the $^{13}$C NMR spin-lattice relaxation rate
varies approximately as $T^3$.

\begin{figure}[htb]
\centering
\includegraphics[height=8cm]{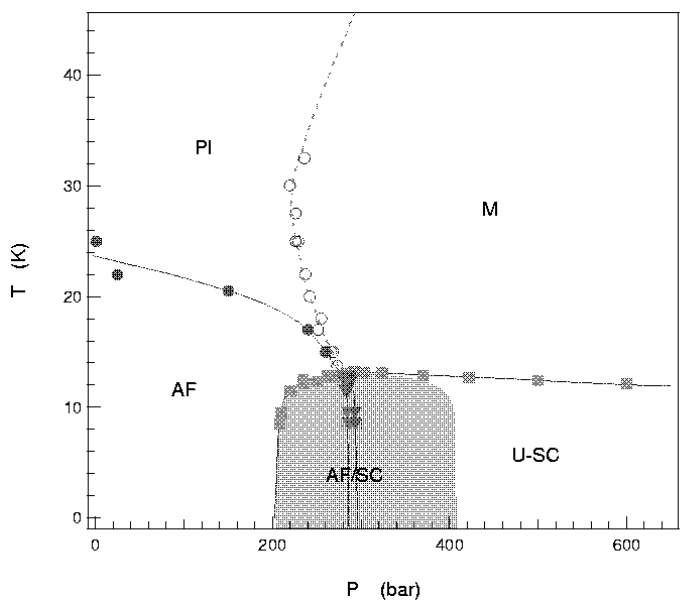}
\label{}
\caption{Temperature versus pressure phase diagram
of $\kappa$-(BEDT-TTF)$_2$Cu[N(CN)$_2$]Cl~\cite{french}. The
antiferromagnetic (AF) critical  line $T_N(P)$ (dark circles) was
determined from  NMR relaxation rate while   $T_c(P)$  for  unconventional
superconductivity (U-SC: squares) and the metal-(Mott) insulator
$T_{MI}(P)$ (MI: open circles) lines were  obtained from the  AC
susceptibility. The  AF-SC boundary (double dashed line)
separates two regions
of inhomogeneous phase coexistence (shaded area).}
\label{lefphase}
\end{figure}

The most recent NMR studies are also the most detailed.
Lefebvre {\it et al.} carried out NMR
and simultaneous ac susceptibility studies of
single crystals of $\kappa$-(BEDT-TTF)$_2$Cu[N(CN)$_2$]Cl
within a hydrostatic helium gas-pressure cell~\cite{french}.
The phase diagram deduced is shown as Figure~\ref{lefphase}.

As has been mentioned above,
at low pressures, this salt is an insulator.
Below 25~K, the magnetic moments of the conduction
holes become ordered into an antiferromagnetic (AF) state;
in other words, alternate holes, each localised
on a dimer (pair of molecules- see Figure~\ref{kappapic}),
have opposite spin.
As the temperature rises, the magnetic order
disappears, and the system is a paramagnetic insulator (PI).

On increasing the pressure, the high-temperature phase of
$\kappa$-(BEDT-TTF)$_2$Cu[N(CN)$_2$]Cl
becomes metallic (M); in other words the holes
are free to move about the crystal, conducting electricity.
At low temperatures, the increases of pressure result
in unconventional superconductivity (U-SC).
Over a restricted region (shaded),
there is an inhomogeneous phase within which
superconductivity coexists
with regions of antiferromagnetism.

The intersection of the metal-insulator (MI) boundary with
the Neel temperature, $T_{\rm N}(P)$, at  ($P^*, T^*)$
is of great interest, since it
shows the absence of a boundary between the
metallic and antiferromagnetic (AF) phases.
This confirms suggestions about the
absence of
itinerant  antiferromagnetism in $\kappa-$phase BEDT-TTF salts~\cite{ross}
and the relevance of a description of magnetic ordering
in terms of interacting spins localized on dimers~\cite{ross,alsoran}.
In other words, the antiferromagnetic interactions are associated
solely with the bands derived from the BEDT-TTF molecules,
and not in any way related to the Cu present in the anions~\cite{schmalian}.
The fact that superconducting (SC) and AF phases
overlap  below $P^*$ indicates that
superconductivity can be directly stabilized from the insulating phase,
a phenomenon noted in studies of other BEDT-TTF superconductors~\cite{lubcz}.

\begin{figure}[htb]
\centering
\includegraphics[height=8cm]{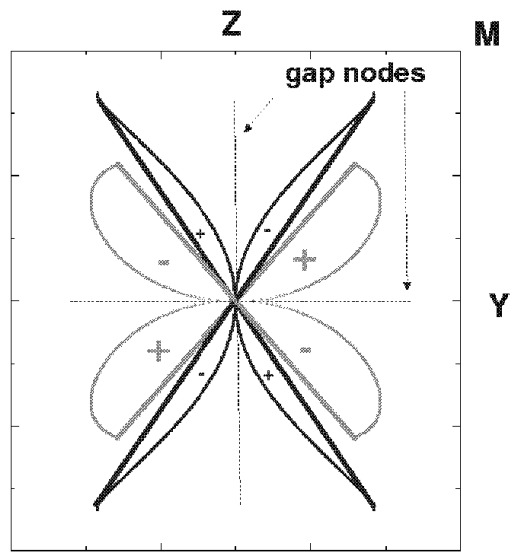}
\caption{Polar plot of the superconducting gap functions of the
two bands (i.e. quasi-one-dimensional and quasi-two-dimensional) at the Fermi surface of
$\kappa$-(BEDT-TTF)$_2$Cu(NCS)$_2$,
calculated by Schmalian~\cite{schmalian}.
The amplitude of the gap corresponds to the distance from the
origin. Gap nodes occur along the $y$ and $z$ directions;
additional suppression of the gap along the diagonal is
caused by the gap between quasi-one-dimensional and quasi-two-dimensional Fermi-surface
(compare Figure~5(a)).}
\label{schmgap}
\end{figure}

Further support for the importance of antiferromagnetic fluctuations
has emerged from theoretical calculations which are able to predict the
behaviour of $1/T_1$ as a function of temperature~\cite{charffi}.
Similarly, using a two-band description of the antibonding orbitals
on a BEDT-TTF dimer and and intermediate local Coulomb repulsion
between two holes on one dimer, Schmalian~\cite{schmalian} was able to
determine the magnetic interaction and superconducting gap
functions of a typical $\kappa$-phase BEDT-TTF
salt within the fluctuation-exchange approximation
(see also References~\cite{alsoran,aoki}).
Schmalian~\cite{schmalian} found that the pairing interaction within this
model is determined by interband coupling,
additionally affected by spin excitations of the quasi-one-dimensional
Fermi-surface sections; moreover, he
was able to predict superconducting transition
temperatures $T_{\rm c} \approx 10$~K, close to those observed
experimentally. The predictions for the shape of the order
parameter within Schmalian's model are shown in Figure~\ref{schmgap}.
\subsubsection{Direct measurements of the anisotropic order parameter.}
\begin{figure}[htb]
\centering
\includegraphics[height=8cm]{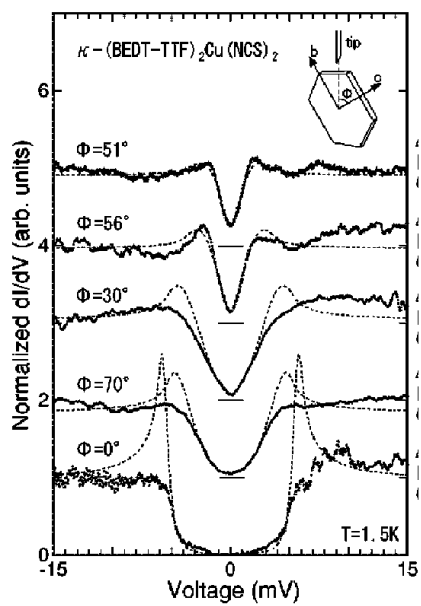}
\caption{$({\rm d}I/{\rm d}V)$ data obtained in tunnelling experiments
on the lateral surfaces of a \cuscn ~single crystal (solid lines; $T=1.5$~K).
The inset shows the definition of the angle $\phi$ used to denote
the tunnelling direction.
The dashed curves are a fit to a d-wave gap model
(after Reference~\cite{araiprb}).}
\label{arai}
\end{figure}

Direct evidence for the anisotropic nature of the superconducting
order parameter in \cuscn ~has come from scanning-tunnelling microscopy
(STM) studies of single crystals at low temperatures~\cite{araiprb}.
Figure~\ref{arai} shows typical data; it is fairly obvious from the raw
data that the $({\rm d}I/{\rm d}V)$ curves are quite strongly
affected by the in-plane
tunnelling direction (defined in the Figure by the angle $\phi$).
Such data have been fitted to d-wave gap models with some success,
yielding maximum gap values $2 \Delta_0/k_{\rm B}T_{\rm c} \approx 6.7$,
substantially larger than the BCS value of 3.53~\cite{araiprb}.
However, in contrast to the predictions of Schmalian~\cite{schmalian},
the STM data appear to favour nodes directed at $45^{\circ}$ to the
${\bf k_b}$ axis, i.e. rotated by $\pi/4$ compared to those
shown in Figure~\ref{schmgap}.

This contradiction seems to have been resolved
very recently
by new calculations of Kuroki {\it et al.},
who showed that the orientation of the calculated node pattern was rather
critically dependent on the fine details of the bandstructure~\cite{kuroki}.
By choosing suitable parameters, Kuroki {\it et al.} were able to
reproduce the node orientation observed in the STM experiments.\footnote{In
this context it should be noted that the transfer integrals used by
Schmalian as input parameters for his calculation~\cite{schmalian}
differ somewhat from more recent, accurate values~\cite{goddard}.}

An attempt has also been made to observe the gap nodes in
\cuscn ~using GHz conductivity measurements~\cite{marijesc}.
However, although these data also suggest an anisotropic gap,
their interpretation remains controversial~\cite{hillcomms}.
\subsubsection{Penetration-depth, Muon Spin Rotation and thermal conductivity measurements.}
More recent
penetration depth measurements in the superconducting state
of the $\kappa$-phase BEDT-TTF salts
also seem to support a d-wave picture.
Early studies~\cite{harshman,lang} were complicated by
difficulties associated with vortex dynamics
(see References~\cite{musr,carrington,tsu} for a discussion);
in more recent work~\cite{carrington,tsu,pinteric}, the importance of
neutralising the influence of external fields has been
realised, and very large values of the penetration length
for magnetic fields in the highly-conducting planes
of $\lambda_{\perp} \approx 100$~microns~\cite{carrington,pinteric}
and $\lambda_{\perp} \sim 0.2$~mm~\cite{tsu}
have been found;
millimetre-wave studies~\cite{jpr,penetrate} are
in broad agreement with these findings.
In all of these studies, the
penetration depth shows a non-BCS-like behaviour
as a function of $T$,
leading to statements such as ``our measurements give strong
evidence for... low-lying excitations...;
values of $\lambda_{||}$ are significantly closer
to those required for d-wave superconductivity''~\cite{carrington}
and ``magnetic field penetration depth results...
indicate an anisotropic superconductivity of a gapless
nature''~\cite{tsu}.

Initial muon-spin rotation ($\mu$SR)~\cite{stevereview}
studies of $\kappa$-phase BEDT-TTF
salts gave inconclusive results because of the complex vortex dynamics;
Reference~\cite{musr} shows that earlier, apparently
definitive statements about the nature of the superconductivity
in BEDT-TTF salts are based on over-simplified analysis.
Subsequently, more sophisticated analysis of
the temperature dependence of $\mu$SR data appears to favour
a superconducting gap with nodes~\cite{pratt}.

The presence of nodes is also suggested by measurements of the
thermal conductivity~\cite{belin}, which
is proportional to $T$ below the superconducting
critical temperature $T_{\rm c}$.
Finally, it has been remarked that the sensitivity
of $T_{\rm c}$ to disorder in the $\kappa$-phase BEDT-TTF
salts points to a non-BCS mechanism for
superconductivity~\cite{xsu}.
\subsection{The $\kappa$-phase BEDT-TTF salts; evidence against gap nodes.}
\begin{figure}[htbp]
\centering
\includegraphics[height=8cm]{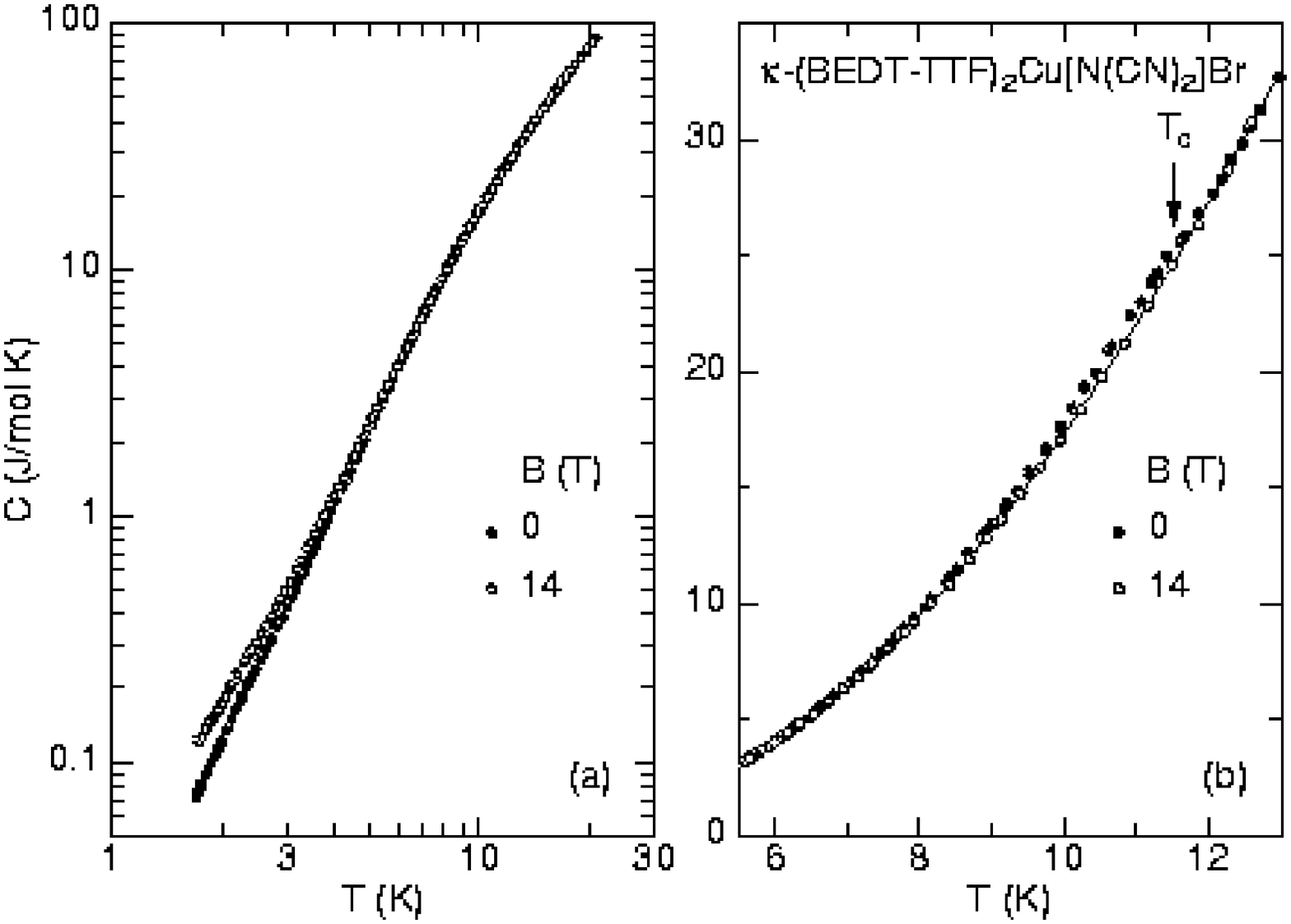}
\vspace{10mm}
\includegraphics[height=8cm]{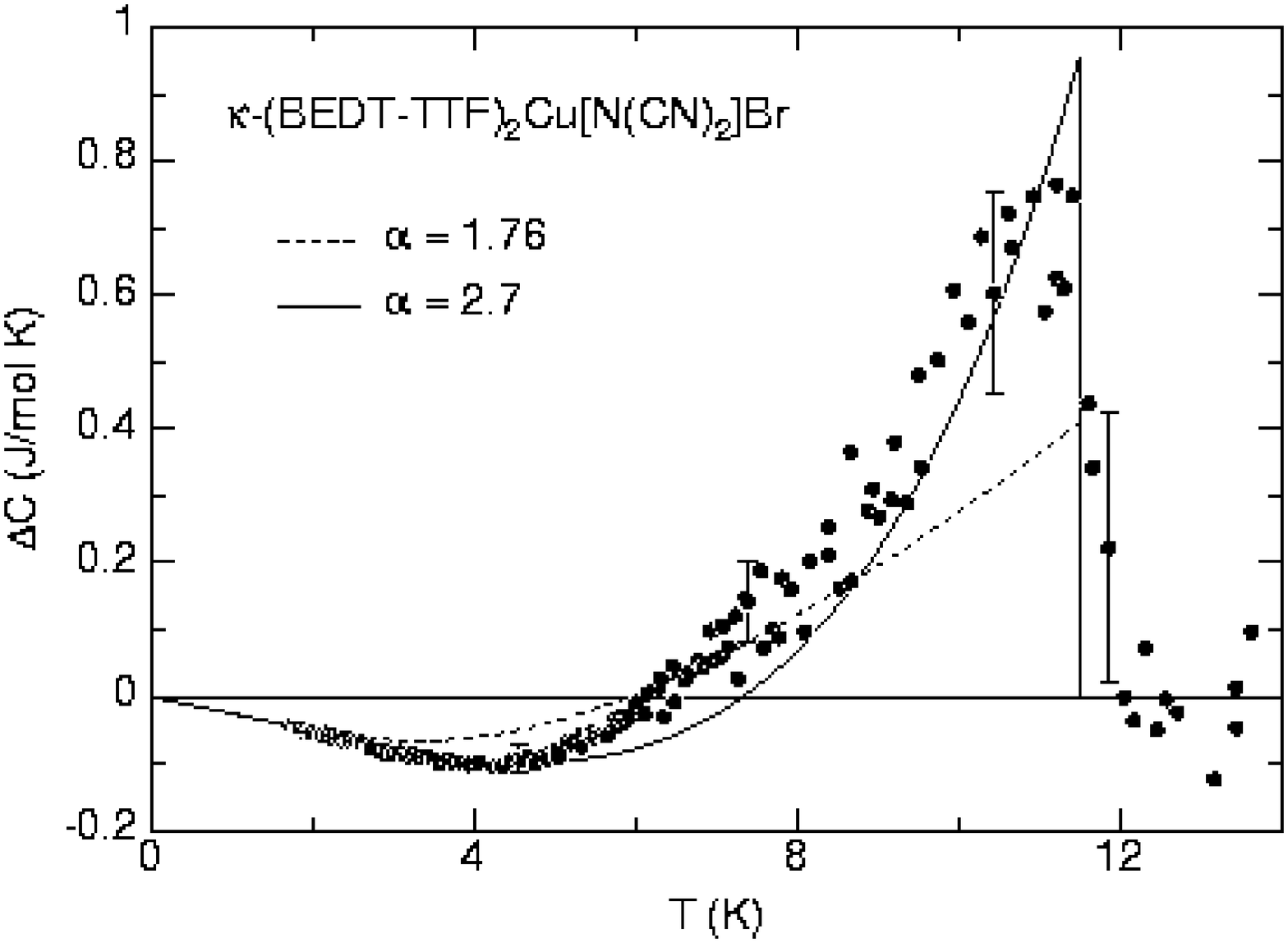}
\caption{Temperature dependence of the specific heat of
$\kappa$-(BEDT-TTF)$_2$Cu[N(CN)$_2$]Br  in the superconducting
($B = 0$) and normal ($B = 14$\,T) state shown (left) for the
complete temperature range and (right) for the region close to
$T_c = 11.5$\,K. The solid line in the centre figure is a polynomial fit
to the 14\,T data. Bottom: specific-heat difference between the
superconducting ($B=0$)
and normal state ($B=14$~T) with theoretical BCS curves for weak (dashed line) and strong
(solid line) coupling (after Reference~\cite{elsinger}.}
\label{elsie}
\end{figure}

The chief evidence against gap nodes has come from heat capacity
measurements by the group of Wosnitza~\cite{elsinger}.
Typical data are shown in Figure~\ref{elsie} for $\kappa$-(BEDT-TTF)$_2$Cu[N(CN)$_2$]Br;
the experiments compare the heat capacity measured in zero magnetic field (superconducting) and
at $B=14$~T (normal). The difference between the
measured heat capacities at $B=0$ and $B=14$~T
(Figure~\ref{elsie}, right-hand side), attributed to the superconductivity,
appears to follow a BCS, strong-coupling-like temperature dependence,
implying a fully-gapped superconducting state.

It is interesting to speculate as to why the heat capacity data should imply
such a different superconducting groundstate from the other experimental probes.
Discussion on this matter has encompassed experimental difficulties,
including the possible field dependence of the 
thermometry used~\cite{neilstuff}, the extreme difficulty in subtracting
the other contributions to the heat capacity,
the presence of other field-dependent contributions to the
heat capacity~\cite{neilstuff,maki1} and the
evolution of the form of the superconducting
order parameter with magnetic field~\cite{newcharfi}.
At present, the matter is unresolved.
\subsection{The isotope effect.}
{\it The isotope effect}~\cite{ashcroft} was a key piece of evidence
for the BCS (phonon-mediated) model of superconductivity. It was found
that the superconducting critical temperature of a
particular isotope of an elemental superconductor
was often proportional to $M^{-\frac{1}{2}}$,
where $M$ is the isotopic mass.
This is a strong indication that phonons are
providing the basic energy scale for the superconductivity;
if phonons are regarded as oscillations of an array of
masses (the atoms or molecules)
on ``springs'' (the chemical bonds-
see References~\cite{singlebook,ashcroft}), then the
characteristic phonon energy should scale as
$M^{-\frac{1}{2}}$.
The absence of an isotope effect or the presence
of an anomalous isotope effect are suggestive of
a non-BCS-like mechanism for superconductivity.

Workers at Argonne National Laboratories have made very careful
studies of the isotope effect. These show only very small
influences of the molecular mass on $T_{\rm c}$~\cite{isotope};
in some cases (replacement of
the 8 $^1$H in BEDT-TTF with $^2$H)
a ``negative isotope effect'' was observed, whereas
a very small shift of $T_{\rm c}$ in the expected direction
was seen when S and C atoms were substituted with heavier
isotopes~\cite{isotope}.
These data could signal support for {\it either}
phonon-mediated or more exotic forms of superconductivity~\cite{isotope}.
One of the effects of deuteration may be to
change the length of the C-H bonds on the ends of the BEDT-TTF
molecule~\cite{isotope}. Therefore, deuteration is analogous
to the effects of anisotropic expansion~\cite{stress} and/or pressure;
both the phonon system and the electronic
system (see Sections~\ref{s2p5} and \ref{s3p1p1})
are likely to be affected~\cite{klehe}.
This matter at present therefore remains inconclusive.
\setcounter{footnote}{1}

\subsection{Raman scattering and infrared measurements; the role of phonons}
\label{s3p4}
\subsubsection{Experimental data and interpretation}
Raman scattering\footnote{An introduction to Raman
scattering at a good, basic level
is given in {\it Optical properties of solids},
by Mark Fox (Oxford University Press, 2001).}
and infrared reflectivity have been used extensively
in studying quasi-two-dimensional organic superconductors.
There are several good motivations for
examining the infrared reflectivity;
(i)~reflectivity can potentially probe low-energy excitations which are
characteristic of the bare, undressed, band electrons~\cite{quader} (see
Sections~\ref{s2p4} and \ref{s2p5});
(ii)~models (e.g. Reference~\cite{rice}) can be used
to obtain an indication of phonon-specific
electron-phonon
interactions~\cite{olga} from reflectivity data;
(iii)~the mid-infrared reflectivity of organic molecular metals
exhibits a large `hump', which has been interpreted as a direct measure of
the Coulomb correlation energy~\cite{mazumdar}.
Thus, infrared studies potentially enable effects due to the bare
bandstructure, the electron-phonon interactions and the
electron-electron interactions to be
distinguished~\cite{olga,rozenberg}.

\begin{figure}[htbp]
\centering
\includegraphics[height=8cm]{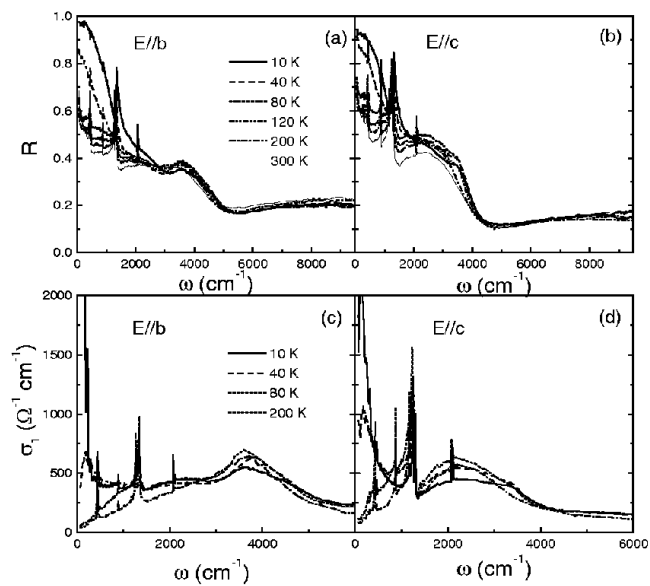}
\caption{Reflectivity of \cuscn ~as a function of photon
energy for infrared polarised
parallel {\bf b} (a) and {\bf c} (b) respectively.
Data for several temperatures are shown. (c) and (d) show the
corresponding frequency-dependent conductivity, $\sigma (\omega)$
(after Reference~\cite{wang}).
Note the sharp peaks due to phonons and the
increase in low-frequency
conductivity (i.e. metallic behaviour)
as the temperature decreases.}
\label{wang1}
\end{figure}

\begin{figure}[htbp]
\centering
\includegraphics[height=8cm]{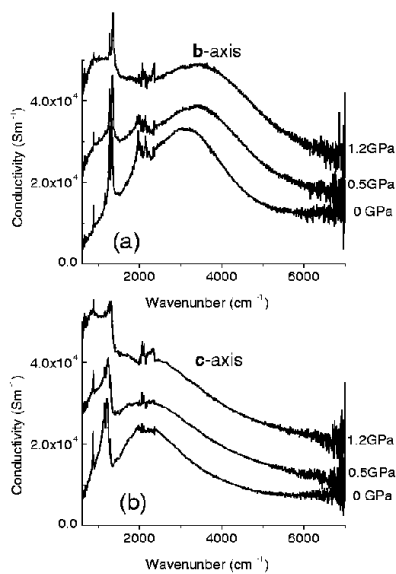}
\caption{Room temperature
frequency-dependent conductivity $\sigma (\omega)$
of \cuscn ~for
infrared polarised parallel to {\bf b} (top) and {\bf c}
(bottom) for pressures 0, 0.5 and 1.2~GPa
(after Reference~\cite{klehe}).
Note the increase in low-frequency conductivity
as the pressure increases.}
\label{klehe1}
\end{figure}

Figures~\ref{wang1} and \ref{klehe1} show typical data,
either as raw reflectivity (Figure~\ref{wang1}(a) and (b))
or as conductivity derived from the reflectivity~\cite{klehe,wang}.
Note the presence of the broad `hump' around 3000~cm$^{-1}$
mentioned above, the sharp lines due to phonons and the
low-frequency conductivity, which increases as the temperature
is lowered (Figure~\ref{wang1}(c), (d)) and/or the
pressure is raised (Figure~\ref{klehe1}).
The latter feature is interpreted as a Drude~\cite{ashcroft} peak (see
References~\cite{klehe,wang} and references therein),
which becomes more prominent as the material's metallic
character increases with increasing pressure or decreasing 
temperature.

The interpretation of reflectivity data
from the $\kappa$-phase BEDT-TTF salts is still
somewhat varied. For example, Wang {\it et al.}
argue quite convincingly
that data such as those in Figure~\ref{wang1} are
best understood in terms of polaron absorption
(see Reference~\cite{wang} and references therein).
The broad hump, most prominent at high temperatures,
is interpreted as
photon-assisted hopping of small polarons (c.f. Reference~\cite{mazumdar});
the sharp Drude peak that develops at low temperature
together with spectral weight in the mid infrared
are attributed to coherent and incoherent bands of small
polarons. Thus, Wang {\it et al.} associate the
transition from insulating-like
behaviour at high temperature to metallic behaviour at low temperature
(Section~\ref{s2p6}) with a crossover from localised small polarons to
coherent large polarons~\cite{wang}. 

Klehe {\it et al.} interpret their pressure-dependent
reflectivity differently, but were able to
extract optical
quasiparticle masses by integrating over the whole
of the measured optical conductivity~\cite{klehe}.
The measured optical mass shows a small linear
pressure dependence (Figure~12), in contrast to the
rapidly-varying effective mass $m^*$ shown in Figure~11.
As the optical mass is closely related to the
bare band mass (see Sections~\ref{s2p4} and \ref{s2p5}), this implies that
the rapid variation of $m^*$ with pressure
is associated with significant changes
in the interactions which renormalise the
quasiparticle mass~\cite{klehe}.
By comparing their reflectivity data with
Raman experiments as a function of pressure~\cite{klehe2,klehe3},
Klehe {\it et al.} suggest that it is the electron-electron
interactions which are most important in determining the
pressure dependence of the superconductivity.

A number of Raman studies of $\kappa$-phase BEDT-TTF
salts are reviewed and compared with those on
high $T_{\rm c}$ cuprates in Reference~\cite{Faulques},
which also reports a microRaman study of low-frequency
(energies less than $\sim 100$~cm$^{-1}$) phonon modes
in $\kappa$-(BEDT-TTF)$_2$Cu[N(CN)$_2$]Br.
This particular paper is interesting because phonon
self-energy effects are monitored as the salt
enters its superconducting phase;
the resulting energy shifts are interpreted in terms
of a strong-coupling BCS model (i.e. fully-gapped),
yielding an electron-phonon
coupling constant $\lambda\approx
0.97$~\cite{Faulques}.
Analogous Raman data involving higher-frequency
phonons ($\sim 500-1500$~cm$^{-1}$)
were reported by Eldridge {\it et al.}, who observed
a shift of surprisingly high-energy phonons (i.e. energies much
greater than $k_{\rm B}T_{\rm c}$)
at $T_{\rm c}$~\cite{eld4}.
Eldridge {\it et al.} speculated that the
``normal electron-phonon interaction may not be responsible
for the frequency change, but that the superconducting
transition may involve a change in either the geometry
or the arrangement of the BEDT-TTF molecules to which
the particular mode is sensitive~\cite{eld4}.
(Reference~\cite{eld4} is also notable as a good survey of
earlier Raman studies.) 

\begin{figure}[htbp]
\centering
\includegraphics[height=8cm]{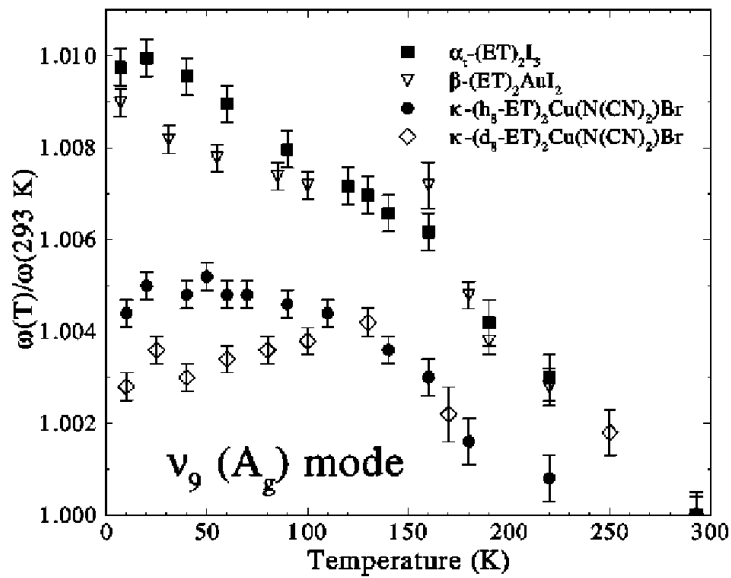}
\caption{Normalised frequency of the $\nu_9$ ($A_g$) mode
in $\alpha-$(BEDT-TTF)$_2$I$_3$,
$\beta$-(BEDT-TTF)$_2$AuI$_2$,
$\kappa$-(BEDT-TTF)$_2$Cu[N(CN)$_2$]Br
and $\kappa$-(d8-BEDT-TTF)$_2$Cu[N(CN)$_2$]Br
as a function of temperature;
in the latter material, the eight
terminal hydrogens have been replaced with
deuterium. 
For $\alpha-$(BEDT-TTF)$_2$I$_3$ and
$\beta$-(BEDT-TTF)$_2$AuI$_2$, the usual
hardening (increase in frequency) of the mode
occurs as the lattice contracts with decreasing temperature.
However, in the case of the two $\kappa$-phase salts,
softening (reduction in frequency) occurs below
the temperature at which antiferromagnetic fluctuations
become important (after Reference~\cite{eld3}).
}
\label{eldfig}
\end{figure}

Finally, infrared and Raman measurements also show
evidence for the coupling
of antiferromagnetic
fluctuations to phonons
which are suspected of involvement in the
superconductivity~\cite{eldridge,eld3}.
Some representative data are shown in Figure~\ref{eldfig},
which shows the normalised frequency of the $\nu_9$ ($A_g$) mode
in $\alpha-$(BEDT-TTF)$_2$I$_3$,
$\beta$-(BEDT-TTF)$_2$AuI$_2$,
$\kappa$-(BEDT-TTF)$_2$Cu[N(CN)$_2$]Br
and $\kappa$-(d8-BEDT-TTF)$_2$Cu[N(CN)$_2$]Br
as a function of temperature;
in the latter material, the eight
terminal hydrogens have been replaced with
deuterium. 
For $\alpha-$(BEDT-TTF)$_2$I$_3$ and
$\beta$-(BEDT-TTF)$_2$AuI$_2$, the usual
hardening (increase in frequency) of the mode
occurs as the lattice contracts with decreasing temperature.
However, in the case of the two $\kappa$-phase salts,
softening (reduction in frequency) occurs below
the temperature at which antiferromagnetic fluctuations,
measured in NMR studies~\cite{eld3},
become important~\cite{eld3}.
Note that $\alpha-$(BEDT-TTF)$_2$I$_3$
has no antiferromagnetic fluctuations present~\cite{eld3}.
\subsubsection{Summary of optical data}
To summarise the optical data, it must be said that
the picture emerging from
infrared and Raman experiments is still unclear.
The interaction between the phonon system
and antiferromagnetic fluctuations does seem to be
important for superconductivity in
the $\kappa$-phase BEDT-TTF
salts~\cite{eld3}. However, whereas some interpret their data
in terms of a strong-coupling, fully-gapped BCS
picture, with an electron-phonon
coupling constant $\lambda \sim 1$~\cite{Faulques},
others are able to use
optical data and ab-initio
calculations (see {\it e.g.} References~\cite{ko,vlasova1} and references
therein for a review and discussion) to derive values
$\lambda \sim 0.3$.
Similarly, infrared measurements~\cite{visentini}
have been taken to suggest that ``{\it intra}molecular vibrations are not
directly involved in the superconductivity
mechanism'' and that the electron-phonon coupling constant
is small; others
found ``no evidence that higher values of $T_{\rm c}$
are achieved through a softer lattice'' ({\it i.e.} that
{\it inter}molecular vibrations are not involved in a simple, BCS-like
manner) and that the superconductivity ``is likely to be
a much more complicated situation than that described by
... simple BCS''~\cite{dressel}. The latter view
is in part supported by the interpretation of pressure-dependent
infrared and Raman data~\cite{klehe,klehe2,klehe3},
which seems to suggest a dominant role for electron-electron interactions.

These divergent opinions may well reflect as-yet inadequate
interpretative tools (see e.g. Reference~\cite{visentini7}
for a discussion). In spite of this disagreement, it is clear
that the optical measurements do suggest important roles for phonons,
antiferromagnetic fluctuations and/or other electron-electron interactions
in the superconductivity of $\kappa$-phase BEDT-TTF salts.

\subsection{The role of effective dimensionality.}
We have seen above that antiferromagnetic
spin fluctuations are not
the only important ingredient in superconductivity in the
$\kappa$-phase BEDT-TTF salts; we also
described good evidence that the
phonon and magnetic systems interact, perhaps
suggesting that the superconducting mechanism
in BEDT-TTF salts may involve both phonons
and spin fluctuations~\cite{caulfield1,lee,eldridge,eldr}.
In this context, it is interesting to contrast
the behaviour of the $\kappa$-phase BEDT-TTF salts
with the isostructural superconductor
(MDT-TTF)$_2$AuI$_2$ ($T_{\rm c}=4.2$~K)~\cite{mdt}.
In NMR studies of (MDT-TTF)$_2$AuI$_2$, a Hebel-Slichter
peak suggestive of conventional BCS-type
superconductivity was observed~\cite{mdt2};
furthermore, the specific heat capacity of this material
was unambiguously BCS-like~\cite{mdt}.

A possible explanation for this difference
comes from the anisotropy of the upper critical
field in (MDT-TTF)$_2$AuI$_2$,
which shows~\cite{mdt} that the
interlayer coupling is much larger than in
$\kappa$-phase BEDT-TTF superconductors.
Models for superconducting pairing
mediated by antiferromagnetic
fluctuations are sensitive to the degree
to which the Fermi-surface may nest~\cite{aoki,schmalian,alsoran};
the increased ``three dimensionality'' of
(MDT-TTF)$_2$AuI$_2$ may be enough to ensure that
required degree of nesting is not present, ruling out this mechanism
of superconductivity. It is interesting to speculate
about a systematic study of isostructural
superconductors with different cations, in which
a transition from BCS-like superconductivity
to a less conventional mechanism could be engineered.
\subsection{The $\beta$-phase BEDT-TTF salts.}
\label{s3p4p2}
In many respects the behaviour of the $\beta$-phase
BEDT-TTF salts is somewhat similar to that of the
$\kappa$-phase salts. For example,
NMR studies indicate
the lack of a Hebel-Slichter peak, suggesting non-BCS-like
superconductivity~\cite{ishi5134}.
Support for a more exotic species of superconductivity
also comes from tunnelling experiments, which
give maximum energy gap values for $\beta$-(BEDT-TTF)$_2$AuI$_2$
which are almost four times larger than expected from BCS theory~\cite{ishi5145}.
Nevertheless, as in the case of the $\kappa$-phase salts,
recent calculations for $\beta$-phase salts,
based on optical data~\cite{girlando},
also suggest an important role for phonons in
the mechanism for superconductivity.

There may be one important difference between
the $\kappa$ and $\beta$-phase salts.
Measurements of the electronic susceptibility as a function of pressure
in the latter materials seem to show that a variation in the density
of states is not a major influence on the superconducting $T_{\rm c}$~\cite{ishi5133}
(c.f. the data described in Section~\ref{s2p5}).
\subsection{The nature of superconductivity in organics; a summary}
In summary, the following remarks may be made.
\begin{itemize}
\item
The greater proportion of experimental techniques
(NMR, tunnelling, $\mu$SR, penetration-depth measurements,
thermal conductivity) support a non-BCS (i.e. unconventional)
mechanism for superconductivity in the organics.
\item
The data suggest a d-wave order parameter arrangement,
with nodes directed along the $x$ and $y$ intraplane directions.
NMR studies suggest the involvement of antiferromagnetic fluctuations.
\item
Theoretical calculations invoking the involvement
of antiferromagnetic fluctuations are able to reproduce
the node pattern and simulate superconducting transition
temperatures of a reasonable size.
\item
The interpretation of isotope effect and heat capacity data
is unclear at the moment.
Differences in interpretation of reflectivity
and Raman data have also resulted in apparently
conflicting statements about the nature of the
superconductivity in the organics,
whilst nevertheless supporting the involvement
of antiferromagnetic fluctuations.
\end{itemize}

\section{The superconducting phase diagram}
\subsection{The broadened transition and the hump.}
\label{s4p1}
\begin{figure}[htbp]
\centering
\includegraphics[height=10cm]{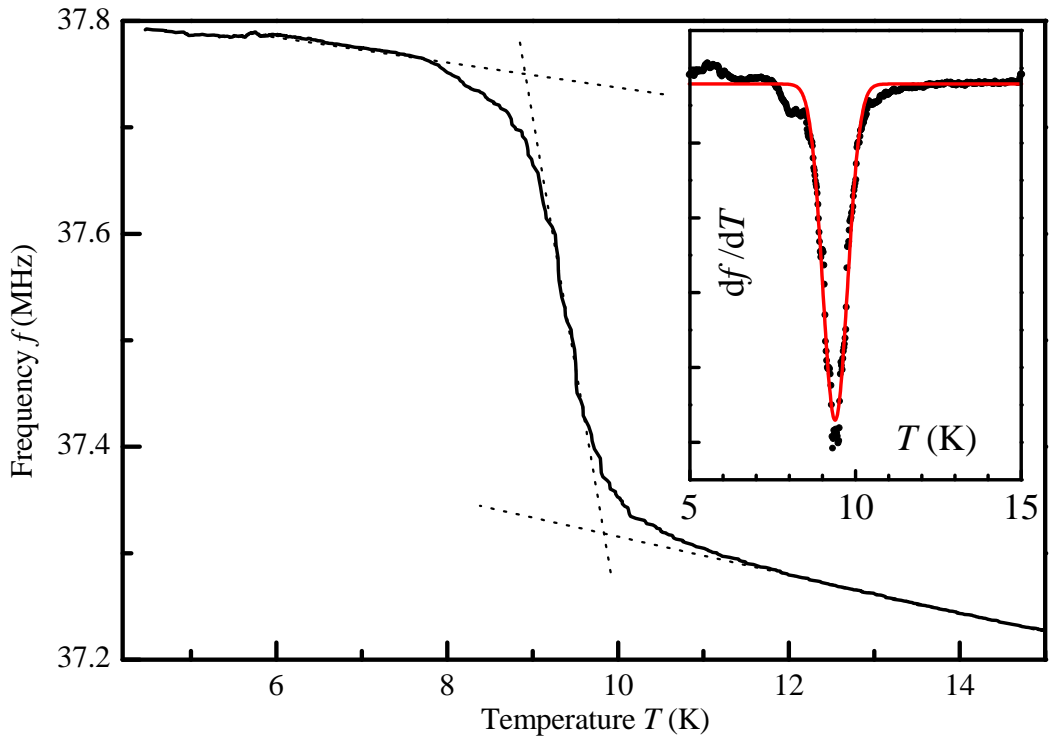}
\caption{MHz penetration data for a single crystal
of \cuscn , shown as resonant frequency $f$ versus
temperature $T$. The superconducting
transition is the steeply sloping region between the more
gentle variations characteristic of superconductivity (low $T$)
and the normal state (high $T$); note
that the complete transition region
occupies a temperature range from
around 8~K to 11~K.
The dotted lines are extrapolations
of the normal-state, transition-region and superconducting-state behaviour.
The intersections of the extrapolations occur at 8.9~T
and 9.8~T, giving $T_{\rm c}^{\rm linear} \approx 9.35$~K (midpoint)
and $\Delta T_{\rm c}^{\rm linear} \approx 0.9$~K.
The inset shows the differential
${\rm d}f/{\rm d}T$ of the data (points) fitted
to a Gaussian (curve) centered on $T_{\rm c}^{\rm Gauss}=9.38$~K,
with a full width of $\Delta T_{\rm c}^{\rm Gauss} \approx 0.7$~K
(after Reference~\cite{neilnew}).
}
\label{chuck}
\end{figure}

In spite of the fact that organic superconductors are
rather clean systems (e.g. the intralayer mean-free path in
a typical sample of \cuscn
is $\sim 2000$~\AA~\cite{goddard}), the transition from normal
to superconducting seems to be rather broad, whatever the method used.
Figure~\ref{chuck} shows the result of a
typical zero-magnetic-field experiment;
in this case a \cuscn ~sample (mean-free-path $\approx 2000$~\AA)
was placed
in a coil forming part of a tank circuit oscillating at $\sim 38$~MHz.
The superconducting to normal transition is seen
as a change in frequency, caused by a change from skin-depth-limited
to penetration-depth-limited coupling of the sample to the MHz
fields~\cite{mielke,neilnew}. The high-temperature
onset of the transition is seen as a deviation
from behaviour characteristic of the normal state close to 11.0~K;
at low temperatures, the frequency starts
to deviate from behaviour characteristic of the superconducting
state just below 8~K, so that the total width of the transition
is around 3~K. Even if one employs more conventional
definitions of the width of the transition~\cite{neilnew}
(see Figure~\ref{chuck}, caption),
the transition is still between 0.7~K and
1~K wide (Figure~\ref{chuck}).
The intrinsic breadth of the transition is almost
certainly responsible
for the wide range of $T_{\rm c}$s quoted for
salts such as \cuscn.\footnote{The resistive
onset often appears to occur around 10.4~K,
and this temperature is regularly quoted
as $T_{\rm c}$; see {\it e.g.} Reference~\cite{janeloff}.
Figure~\ref{chuck} shows that a preferable value
would be around 9.5~K.}

How can the transition be so wide in such a clean material?
The reason is perhaps associated with the exceptionally two-dimensional
nature of the superconductivity.
Let us suppose that there are variations in the local potential
caused by defects etc.. In a three-dimensional superconductor,
the effects of such variations (e.g. on the local density
of states) will be averaged
over a {\it volume} given roughly by $\xi^3$, where $\xi$ is the coherence
length. On the other hand, in a two dimensional superconductor, the
variations will be averaged over an {\it area} $\sim \xi^2$,
corresponding to a much smaller number of unit cells sampled.
This means that the effect of the variations is statistically
much more significant in two-dimensional materials.
Using such an approach, it has been possible to
obtain good estimates of the broadening of the transition
in \cuscn ~\cite{neilnew},
based on a density of spatial variations extracted from
de Haas-van Alphen oscillations.

In a magnetic field, the situation becomes rather worse,
especially in the case of resistive measurements.
Figure~\ref{sdh} has already shown that
there is a ``hump'' in the resistance
between the superconducting and normal behaviour.
This effect is seen over a restricted temperature range
in a number of Cu-containing
$\kappa$-phase salts~\cite{sdh2,lisa,itohump};
it is most noticeable when the current
is driven in the interplane direction~\cite{lisa,itohump}.
A weaker effect can be observed when the current is driven
in-plane, but is suppressed when the number of defects
in the sample
is reduced~\cite{ishihump,comment12}.
By contrast, the ``hump'' is largest
when the current is in the interplane direction for
very pure samples~\cite{zuo1}, so that it may be an
intrinsic feature of these layered materials~\cite{zuo1}.
The hump has been attributed to dissipation due to
superconducting weak links in inhomogeneous samples~\cite{itohump},
magnetoresistance due to a lattice distortion via
coupling with the quantised vortices~\cite{zuo2} and dissipation caused by
fluctuations characteristic of a $d$-wave superconductor~\cite{maki}.
Note that the ``hump'' disappears
in in-plane fields ($\theta=90^{\circ}$)~\cite{zuonew,msnsuper},
indicating that it is associated in some way with the
arrangement of the vortices relative to the
crystal structure~\cite{naughton90deghump}.
This may favour the Josephson-junction model~\cite{ishihump,ambeg},
since this involves a noise voltage associated with thermal
fluctuations which disrupts the phases of the order-parameter
between adjacent Josephson-coupled planes; it will
only be operative when the vortex cores traverse those planes.
It has been already been established that because
$\kappa$-(BEDT-TTF)$_2$Cu(NCS)$_2$ is a very anisotropic
superconductor, the vortex lattice is no longer
a system of rigid rods but consists of a weakly coupled
stack of ``pancake'' vortices, each one confined to
a superconducting plane, with the coupling due to Josephson
or electromagnetic effects~\cite{musr,Mansky}
(see Section~\ref{s4p3}).

\begin{figure}[htbp]
\centering
\includegraphics[height=8cm]{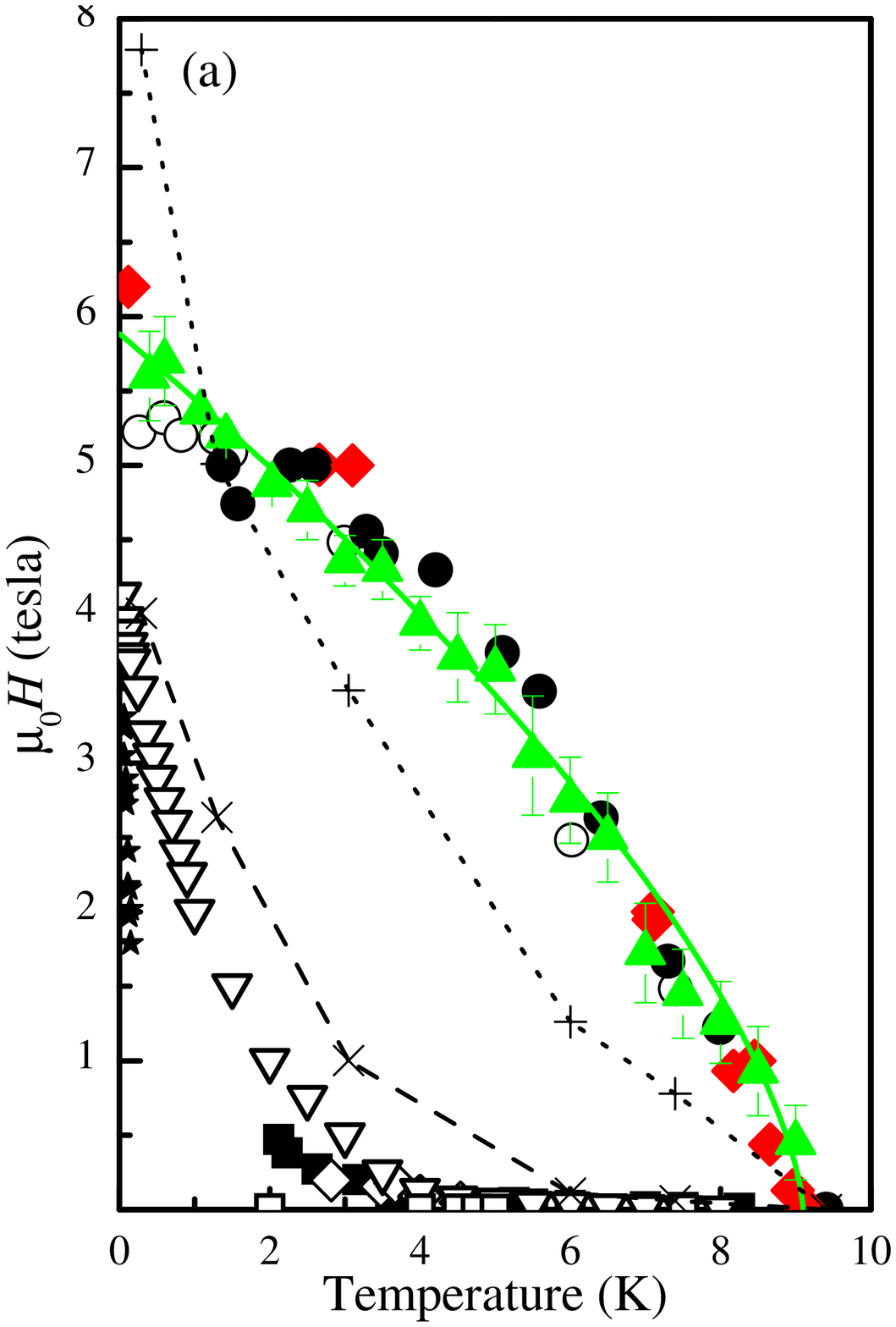}
\includegraphics[height=8cm]{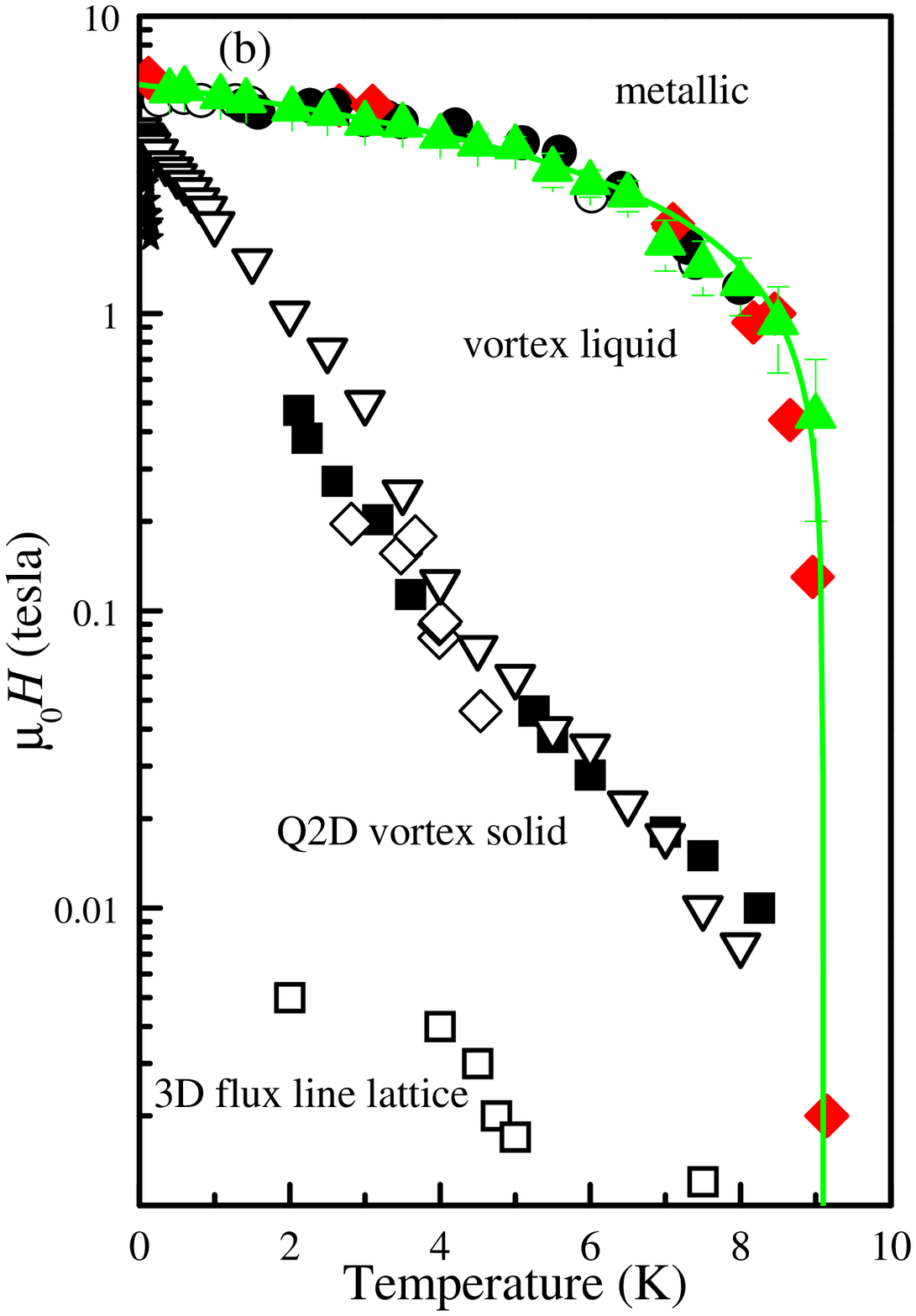}
\caption{Critical fields in \cuscn , plotted on
linear (a) and logarithmic (b)
field scales; {\bf B} is perpendicular to the quasi-two-dimensional layers.
The data for $H_{\rm c2}$ comprise
filled triangles (MHz penetration studies, midpoint)~\cite{jscm},
filled circles (microwave penetration studies)~\cite{belinshib},
open circles (thermal conductivity data)~\cite{belinshib} and shaded
diamonds (magnetisation data)\cite{sasaki,lang}.
The solid curve is proportional to $(T_{\rm c}-T)^{2/3}$,
with $T_{\rm c} =9.1$~K.
The triangles are the irreversibility field from magnetisation;
the filled squares and stars represent two dimensional melting from magnetometry
and GHz studies~\cite{mola} (see also Reference~\cite{mayaffre}).
The hollow squares are from
muon-spin rotation and denote the three dimensional-two dimensional
transition~\cite{musr}.
(a) also shows the low-field (X, joined by dashed lines)
and high-field (crosses, joined by dotted lines)
limits of the resistive transition~\cite{belinshib}.
(After Reference~\cite{mielke}.)}
\label{qpcdemo}
\end{figure}

Whatever the mechanism, these considerations
imply that the hump and the broad transition
from zero to finite resistance are both
phenomena of {\it the mixed state};
{\it i.e.} the dissipative effects mean that resistive measurements
of the critical field are often a poor guide to
reality.
Figure~\ref{qpcdemo}(a) shows a comparison
of the various techniques for deducing 
$H_{\rm c2}$ in \cuscn ~(field applied perpendicular
to the quasi-two-dimensional planes); the data comprise
filled triangles (MHz penetration studies, midpoint),
filled circles (microwave penetration studies),
open circles (thermal conductivity data) and shaded
diamonds (magnetisation data).
(A number of studies have shown that GHz and MHz
penetration measurements are a much less dissipative
probe of the superconducting state than a resistivity measurement;
see Reference~\cite{mielke} and references therein.)
These are compared with the low-field (X, joined by dashed lines)
and high-field (crosses, joined by dotted lines)
limits of the resistive transition~\cite{belinshib}.
Whereas the thermal conductivity, GHz and MHz data are in good
agreement with each other, following quite closely
the convex curve $(T_{\rm c}-T)^{2/3}$~\cite{mielke},
the resistive transition
follows a very different concave temperature dependence,
perhaps related more closely to the irreversibility line than
to $H_{\rm c2}$.

The phase diagrams in Figure~\ref{qpcdemo}(a) and (b) are completed
by the inclusion of further data from $\mu$SR and magnetometry studies.
The triangles are the irreversibility field from magnetisation;
the filled squares and stars represent two dimensional melting from magnetometry
and GHz studies. The hollow squares are from
muon-spin rotation and denote the three dimensional-two dimensional
transition.

\begin{figure}[htbp]
\centering
\includegraphics[height=8cm]{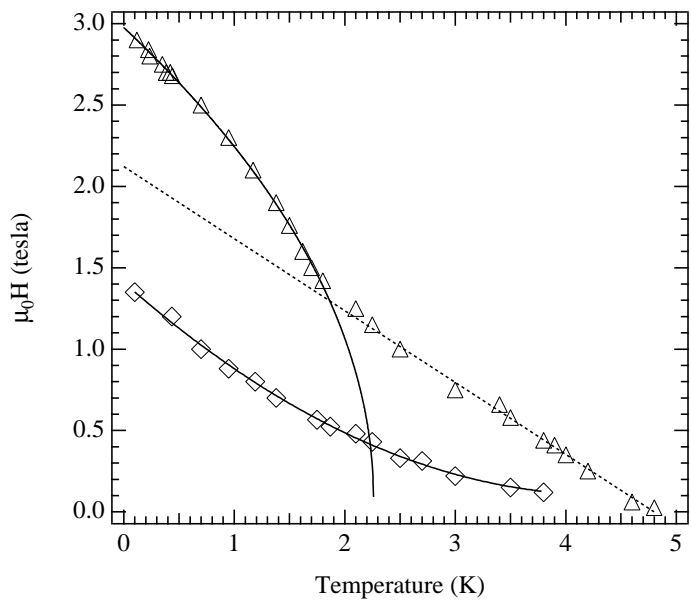}
\caption{Critical fields in \lbets ~(with
{\bf B} perpendicular to the quasi-two-dimensional layers)
measured using MHz penetration techniques.
The data comprise $H_{\rm c2}$ (triangles)
and flux-line lattice melting (diamonds).
The upper solid curve is $H_{\rm c2} \propto (T^*-T)^{1/2}$;
the dashed curve is $H_{\rm c2} \propto (T_{\rm c2}-T)$.
The lower solid curve is a fit of the two-fluid model
expression for the flux-line lattice melting.
(After Reference~\cite{mielke}.)}
\label{betsfig}
\end{figure}

Figure~\ref{betsfig} shows a similar phase diagram for \lbets
~(conducting planes perpendicular to the applied magnetic field),
determined using MHz penetration techniques~\cite{mielke}.
\lbets ~is of great interest because it has a Fermi surface
which is topologically similar to that of \cuscn ,
and very similar effective masses~\cite{mielke}. However,
in contrast to \cuscn , which has an interlayer
transfer integral $t_{\perp} \approx 0.04$~meV,
\lbets ~is rather more three dimensional,
possessing an interlayer transfer integral
$t_{\perp} \approx 0.21$~meV~\cite{mielke}.
In \lbets , $H_{\rm c2}$  has a
linear region $H_{\rm c2} \propto (T_{\rm c}-T)$
that spans from $T_{c}$ to approximately 1.9~K. Below
1.9~K, a definite change in the slope of the upper critical field
occurs, and $H_{\rm c2}$ begins to follow the power
law $H_{\rm c2} \propto (T^*-T)^{\zeta}$, with $T^*$
and $\zeta$
fit parameters; powers
$\zeta$ in the range $0.5-0.7$ provide an adequate fit
to the data.
The behaviour of $H_{\rm c2}$ in \lbets ~at temperatures
below 1.9~K is therefore very similar to that of $H_{\rm c2}$
in \cuscn ~over the whole temperature range shown in
Figure~\ref{qpcdemo}, and is thus characteristic
of a two-dimensional
superconductor with weakly-coupled layers~\cite{mola}.
On the other hand, the linear variation
of $H_{\rm c2}$ in \lbets ~at higher temperatures
follows the expectations of Ginzberg-Landau theory
for three-dimensional superconductors~\cite{tinkham,parks}.
The change in gradient at 1.9~K
is therefore
attributed to dimensional cross-over from quasi-two-dimensional (low temperatures)
to three dimensional (high temperatures)~\cite{mielke}.

Dimensional crossovers with the magnetic field
applied perpendicular to the quasi-two-dimensional planes
have been observed
in artificial quasi-two-dimensional superconducting structures~\cite{white,perpnote}
and in organic superconductors such
as \cuscn~(see Figure~\ref{qpcdemo})\cite{musr,mola};
however, in the majority of these cases, the effect of the crossover
is observed at magnetic fields less than $H_{\rm c2}$.
\lbets ~is perhaps unique in providing the correct
anisotropy for the crossover to be observed in the
behaviour of $H_{\rm c2}(T)$.
It has been mentioned that the interplane
transfer integral in \lbets ~is a factor $\sim 5$ larger than
that in \cuscn . The greater ``three dimensionality'' of
the bandstructure of \lbets ~compared to
\cuscn  ~obviously manifests itself in the
superconducting behaviour (compare Figures~\ref{betsfig}
and \ref{qpcdemo}); whereas \lbets ~exhibits
two dimensional-three dimensional dimensional crossover in its $H_{\rm c2}(T)$ behaviour,
$H_{\rm c2}(T)$ in \cuscn ~is entirely characteristic of a quasi-two-dimensional
superconductor.
\subsection{Anisotropy of the superconducting parameters.}
\label{s4p2}
A number of studies of the anisotropy of the upper
critical field of \cuscn ~have been made, all employing
resistivity measurements~\cite{zuonew,msnsuper,naka}
(for a summary, see Section~3.4 of Reference~\cite{review}).
The angle-dependent behaviour of \cuscn ~seems to be rather
typical of quasi-two-dimensional organic superconductors, and as the
studies of this salt are the most detailed,
we shall focus on them.

As we have seen above, the resistive transition is often a poor
guide to the temperature dependence of $H_{\rm c2}$;
however, most of the anisotropy studies were carried out
at fixed temperatures~\cite{zuonew,msnsuper,naka},
and some attempts were made to choose
a characteristic point in the resistivity which took account
of the fact that the ``hump'' is a feature of the mixed state~\cite{janeloff}.
To the limit of experimental accuracy,
it was found that $H_{\rm c2}$ depended only on $\theta$,
the angle between the normal to the quasi-two-dimensional planes and the magnetic field;
in spite of some expectations to the contrary, it appears
that $H_{\rm c2}$ is rather insensitive to the plane of rotation
of the field~\cite{msnsuper}.

For an exactly in-plane field ($\theta=90^{\circ}$),
$H_{\rm c2}$ reaches its
maximum value of $\mu_0 H_{\rm c2} (T \rightarrow 0) \approx 35$~T in
\cuscn ~\cite{janeloff}.
Away from this orientation, the variation of
$H_{\rm c2}$ with $\theta$ is at first sight
qualitatively similar
to the predictions
of the Ginzberg-Landau
anisotropic effective mass approximation~\cite{zuonew,naka,tinkham,Morris},
\begin{equation}
B_{\rm c2}(\theta) = \frac{B_{\rm c2}(\theta = 0)}
{\sqrt{\cos^{2}(\theta) +  \gamma^{-2} \sin^{2}(\theta)}} ,
\label{gl}
\end{equation}
in which the
superconductivity is destroyed by orbital effects;
here $\gamma$ is the square root of the ratio of the
effective masses for interplane and in-plane motion respectively.
However, whilst Equation~\ref{gl}
has a similar form to the published data~\cite{zuonew,msnsuper,naka},
a very serious failure of this approach
becomes apparent when one compares
the value $\gamma \sim 10$
obtained by fitting $H_{\rm c2}$ data to Equation~\ref{gl}
with the accepted value of
$\gamma \sim 100-350$ obtained
from very careful measurements of the penetration
depths~\cite{carrington,tsu,jpr,Mansky}.
These large values of $\gamma$
occur because the
coherence length perpedicular to the conducting layers,
$\xi_{\parallel}$,
is smaller than the interlayer distance $a$.
The intralayer
overlap of electron wave functions in the superconducting state
is very weak
because $\xi_{\parallel}$ is
shorter than the Josephson tunnelling length
$l_{J} = \eta a$ ($\eta$ is a constant)
between Josephson vortices in
$\kappa$-(BEDT-TTF)$_{2}$Cu(NCS)$_{2}$~\cite{Mansky}.
In sufficiently high magnetic fields parallel to the layers,
flux lines will be trapped
inside the layers;
in such a limit, the compressing effect of the magnetic
field on the Cooper pair wavefunction~\cite{onedimensionalization} exactly
compensates the increasing flux density,
potentially leading to a very high in-plane upper critical
field~\cite{estimate}
if orbital effects are the limiting mechanism.
This is obviously {\it not} observed in the data of
References~\cite{naka,zuonew,msnsuper};
some other mechanism therefore seems to be limiting the
upper critical field close to $\theta=90^{\circ}$.

Kovalev et al.~\cite{kovalev} used an elegant series of
heat capacity measurements to show that
the in-plane $\theta=90^{\circ}$ upper critical field
is in fact limited by processes associated with
the quasiparticle spin in $\kappa$-phase BEDT-TTF
salts.
A possible candidate is the
Pauli paramagnetic limit (PPL), also known as the
Clogston-Chandrasekhar limit~\cite{ccpl1,ccpl2,ccpl3}.
This occurs when the
magnetic energy associated with the spin susceptibility in the
normal state exceeds the condensation energy in the superconducting
state;
for isotropic $s$-wave superconductors it is given by
\begin{equation}
\mu_0H_{\rm PPL}(T=0) = 1.84~T_{\rm c}.
\label{pplformula}
\end{equation}
The PPL mechanism should be roughly isotropic
(as the electron g-factor in organic superconductors
is within a few percent of 2 for all field orientations~\cite{ishiguro}).
Bulaevskii~\cite{bull} considered the case where the paramagnetic limit is
larger than $H_{\rm c2}(\theta = 0)$, but smaller than
$H_{\rm c2}(\theta=90^{\circ})$ determined by orbital effects.
In a similar spirit Nam {\it et al.}
proposed a semi-empirical
description of the critical field
representing
an anisotropic orbital limiting mechanism,
dominant at lower values of $\theta$, combined with an
isotropic PPL-type mechanism which
limits the critical field close to $\theta =90^{\circ}$
(Equation~3 of Reference~\cite{msnsuper});
in practice, with some rearrangement, this equation
gives an identical angle dependence to Equation~\ref{gl}.

A critical comparison between the various
predicted angle-dependences of $H_{\rm c2}$
and resistive critical field data
has been made in References~\cite{zuonew} and~\cite{msnsuper}.
However, none of the formulae give a truly
satisfactory fit over the whole angular range,
the fits in Reference~\cite{zuonew}
describing the data close to $\theta=90^{\circ}$
but apparently failing at angles away from this orientation.
By contrast, the expression used
in Reference~\cite{msnsuper} only
fits the data away from $\theta=90^{\circ}$; data for
$|\theta|$ close to $90^{\circ}$
do not seem to follow the same
dependence.
Two proposals have been made to account for this
difference.
Firstly, Zuo {\it et al.}~\cite{zuonew} argued that
the paramagnetic limiting field could be
calculated from thermodynamic arguments
to be $30 \pm 5$~T, consideably larger
than the value predicted by Equation~\ref{pplformula}.
An alternative proposal was made in Reference~\cite{msnsuper},
based on the inability of conventional formalisms
to fit the angle dependence of $B_{\rm c2}$ close
to $\theta =90^{\circ}$; to explain this, and
the large values of $B_{\rm c2}$, the existence of a
Fulde-Ferrell-Larkin-Ovchinnikov (FFLO)~\cite{FF}
state in high in-plane fields and at low temperatures
was invoked~\cite{msnsuper}.
As we shall see in Section~\ref{s5p1},
the latter explanation seems to be the correct one~\cite{janeloff}.

In summary, the upper critical field $H_{\rm c2}$
in $\kappa$-phase organic superconductors
is limited by orbital effects for most orientations
of the field, but limited by processes involving the
quasiparticle spin when the field lies in the quasi-two-dimensional planes.
This implies that estimates of the interlayer coherence
length derived from the in-plane critical field~\cite{ishiguro}
are incorrect.
A better guide to the true anisotropies of organic
superconductors is derived from magnetisation~\cite{Mansky},
penetration-depth~\cite{carrington} and $\mu$SR studies~\cite{musr,pratt};
such studies indicate $\gamma \approx 350$ for \cuscn .
\subsection{The vortex lattice.}
\label{s4p3}
\begin{figure}[htbp]
\centering
\includegraphics[height=5cm]{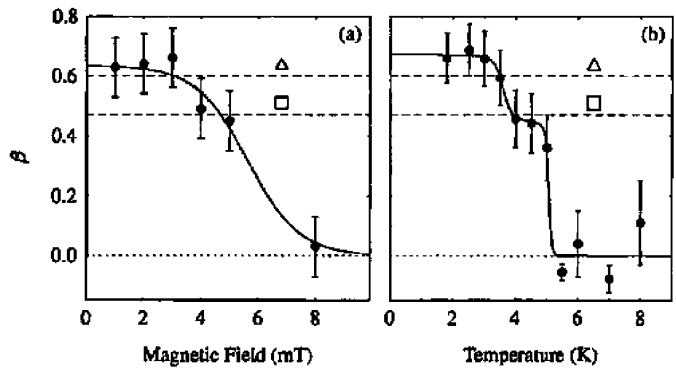}
\caption{Skewness parameter $\beta$ of \cuscn
~plotted as a function of magnetic field
at a temperature of 2~K~(a)
and as a function of temperature
at a field of 3~mT~(b).
The skewness parameter, determined in $\mu$SR experiments~\cite{stevereview},
takes on very characteristic values for different types of flux lattice.
The crossover from three-dimensional flux-line lattice
to decoupled layers is observable as the drop
in $\beta$ seen between 5 and 8~mT in (a).
The expected ranges of $\beta$ for triangular and squre lattices
are denoted by the square and tringle symbols and
the dashed lines. There is some evidence for
a triangular to square transition, especially
in (b) (after Reference~\cite{pratt}).
}
\label{musrfig}
\end{figure}

The vortex lattice in BEDT-TTF superconductors has been studied
by $\mu$SR~\cite{musr,pratt} and by flux decoration techniques~\cite{pratt3}.
Both types of experiment reveal a triangular flux-line lattice
at the lowest fields, with the $\mu$SR data showing some
evidence for a transition to a square lattice at higher
fields and temperatures, just before the layers
become decoupled (see Figure~\ref{musrfig})~\cite{pratt}.

A key question for the $\mu$SR experiments is
the mechanism (Josephson or electromagnetic~\cite{clem})
for the interlayer coupling of the vortices.
Table~\ref{musrtab} summarises some of the recent $\mu$SR data~\cite{pratt},
comparing the experimental magnetic field $B^{\rm meas}_{\rm cr}$ for
the transition from flux-line lattice to decoupled
layers with that calculated assuming electromagnetic coupling
($B^{\rm EM}_{\rm cr}$).
The table also compares measured and calculated temperatures
$T^*$ for the thermal break up of the flux-line lattice.
The comparison between the measured and calculated parameters
allows the coupling mechanism to be inferred.
It appears that coupling in the more anisotropic BEDT-TTF
superconductors is predominantly electromagnetic in origin.
\begin{table}[tbp] \centering
\begin{tabular}{|l|l|l|l|l|l|l|l|l|}
\hline
 Salt & $\lambda_{||}$  & $B^{\rm meas}_{\rm cr}$ & $B^{\rm EM}_{\rm cr}$ &
$\gamma$ & mode & $T_{\rm c}$ & $T^*$ & $T^*_{\rm calc}$ \\
~ & ~ & (mT) & (mT) &  ~ & ~ & (K) & (K) & (K) \\
\hline
$\beta$-(ET)$_2$IBr$_2$ & 0.86 & $>8$ & 2.8 & $<300$ & J & 2.2 & 1.7 & 1.8 \\ \hline
$\kappa$-(ET)$_2$Cu(NCS)$_2$ & 0.54 & 5.9 & 7.1 & $\sim 350$ & EM/J & 9.2 & 3.9 & 4.5 \\ \hline
$\alpha$-(ET)$_2$NH$_4$Hg(SCN)$_4$ & 1.1 & 1.7 & 1.7 & $>500$ & EM & 1.1 & 0.6-0.8 & 1.4 \\ \hline
\end{tabular}
\caption{Summary of recent $\mu$SR data~\cite{pratt}.
The parameters listed are the anisotropy $\gamma$,
the experimental magnetic field $B^{\rm meas}_{\rm cr}$ for
the transition from flux-line lattice to decoupled
layers, the corresponding calculated field assuming electromagnetic coupling
($B^{\rm EM}_{\rm cr}$), the superconducting
transition temperature $T_{\rm c}$
and the measured and calculated temperatures
$T^*$ for the thermal break up of the flux-line lattice.
The comparison between the measured and calculated parameters
allows the coupling mode (Josephson (J) or Electromagnetic (EM))
to be inferred.}
\label{musrtab}
\end{table}
\section{Field-induced superconducting states.}
\subsection{The Fulde-Ferrell-Larkin-Ovchinnikov phase in \cuscn}
\label{s5p1}
There has been great interest in the possibility of the
Fulde-Ferrell-Larkin-Ovchinnikov (FFLO)~\cite{FF}
state in organic superconductors for some time~\cite{Shimahara}.
In a metal in a magnetic field, the
normal quasiparticles have separate spin-up
and spin-down Fermi surfaces
which are displaced due to the Zeeman energy.
In the FFLO state, attractive
interactions of quasiparticles
with opposite spin on
opposite sides of the two Fermi surfaces
lead to the formation of pairs with nonzero total
momentum~\cite{FF,Shimahara} and hence an inhomogeneous
superconducting state.
It was suggested that the quasi-two-dimensional (quasi-two-dimensional)
superconductor $\kappa$-(BEDT-TTF)$_2$Cu(NCS)$_2$
in exactly in-plane magnetic fields is a possible
candidate for the FFLO~\cite{msnsuper,austrians};
in this section, we describe recent
experiments which suggest that this indeed does occur~\cite{janeloff}.
Although the FFLO phase was predicted in the mid 1960s,
the experiments on \cuscn ~seem to be the first
to give definite evidence for such a state
(see Reference~\cite{janeloff} for a critique of
earlier claims).

\begin{figure}[h]
\vspace{20mm}
\begin{center}\leavevmode
\includegraphics[width=0.6\linewidth]{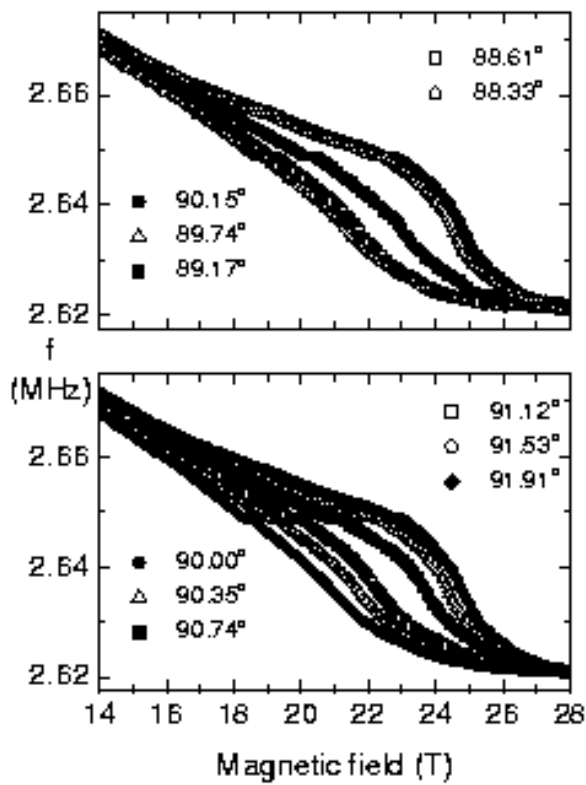}
\caption{TCDS frequency versus magnetic field
for several different values of $\theta$ ($\phi=-45^{\circ}$;
$T=1.39$~K).
The elbow at $B_{\rm L} \sim 22$~T disappears when the
angle differs from $90^{\circ}$ by more than
about $1.5^{\circ}$ (after Reference~\cite{janeloff}).
}\label{fig1}\end{center}\end{figure}

The experiments employed
single crystals of $\kappa$-(BEDT-TTF)$_2$Cu(NCS)$_2$
($\sim 1 \times 0.5 \times 0.1$~mm$^3$; mosaic spread
$< 0.1^{\circ}$) mounted on
or in the coil of a tuned-circuit
differential susceptometer (TCDS)~\cite{janeloff}.
The coil was mounted in a cryostat which
allowed it (and the sample)
to be rotated to all possible orientations
in the magnetic field {\bf B}.
The orientation of the sample
is defined by
the polar angle $\theta$ between
{\bf B} and the normal to the
sample's {\bf bc} planes and the azimuthal
angle $\phi$; $\phi=0$ is
a rotation plane of {\bf B}
containing {\bf b} and the
normal to the {\bf bc} plane.

In order to detect the FFLO,
Reference~\cite{janeloff} examined the {\em rigidity}
of the vortex arrangement, which is predicted
to change on going from mixed state to FFLO~\cite{rigid}.
The sample was mounted with its quasi-two-dimensional planes perpendicular
to the axis of the TCDS coil.
When the quasistatic field
is in the sample planes ($\theta=90^{\circ}$),
the TCDS coil provides
an oscillating magnetic field {\it perpendicular}
to the static field (and the vortices) which
exerts a torque
on the vortices.
The coil in the TCDS forms part of a tank circuit,
so that changes in the rigidity
of the vortices affect the effective ``stiffness'' of the
circuit and therefore shift its resonant frequency $f$.

Figure~\ref{fig1} shows $f$ for
for several angles $\theta=90^{\circ}\pm \Delta \theta$.
Superimposed on the gentle downward trend, due to
the growing flux penetration of the sample,
is an ``elbow'' at $\sim 22$~T for values of
$\Delta \theta$ close to zero.
The elbow indicates
a change in the vortex rigidity which was associated with the
onset of the FFLO state at a magnetic field labelled
$B_{\rm L}$.
The elbow
only occurs for
$|\Delta \theta |$ less than about $1.5^{\circ}$;
this is in good agreement with the calculations
of Refs.~\cite{austrians,rigid}, which predict that the FFLO is only
stable in typical organic
conductors for $|\Delta \theta| < 0.3-2.3^{\circ}$.
For bigger deviations, a substantial
number of closed orbits will be possible
on the Fermi surface~\cite{amro2}, leading to suppression
of the superconductivity due to orbital effects~\cite{janeloff}.

\begin{figure}[h]
\begin{center}\leavevmode
\includegraphics[width=0.5\linewidth]{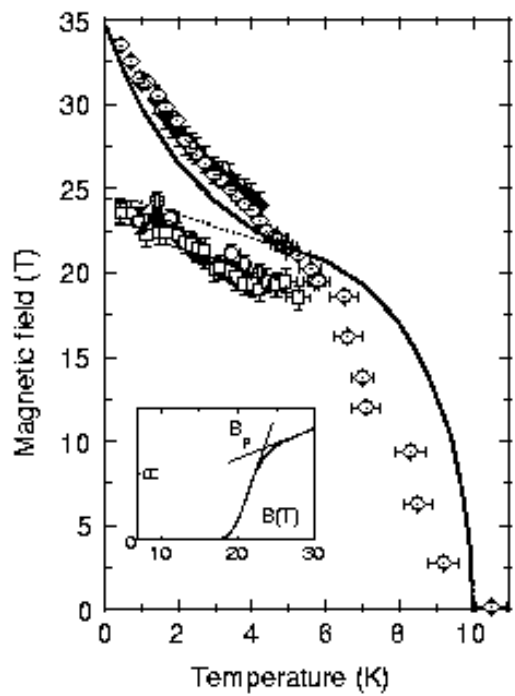}
\caption{Temperature dependence of
the fields $B_{\rm L}$ (squares, circles) and
$B_{\rm p}$ (diamonds; the field $B_{\rm p}$
is defined in the inset as the intersection of
the two extrapolations);
the different symbols indicate data from different samples
and experimental runs.
The data are compared with
the FFLO phase diagram of Shimahara;
the solid curve separates the
superconducting and normal states.
The boundary between the mixed state and
the FFLO is shown as a dotted line (after Reference~\cite{janeloff}).
}\label{fig2}\end{center}\end{figure}

Figure~\ref{fig2} compares $B_{\rm L}$ with
the calculated FFLO phase diagram of Ref.~\cite{shimprb},
derived for a quasi-two-dimensional metal; also plotted are values
of the resistive upper critical field $B_{\rm p}$
($\theta=90.00^{\circ}$),
defined in the inset to Fig.~\ref{fig2}.
The theoretical curves have been
scaled using a $T=0$ $B_{\rm c2}$ of 35~T
and $T_{\rm c}=10$~K.
Even though there are (not unexpected-- see Section~\ref{s4p1})
deviations of $B_{\rm p}$ from
the theoretical dependence of $B_{\rm c2}$,
the data in Fig.~\ref{fig2} bear a
striking similarity to the calculations
of Ref.~\cite{shimprb}.
In particular, $B_{\rm L}$ follows the phase boundary
between the Type-II superconducting state
and the FFLO state (dotted curve) closely, extrapolating
to $B_{\rm p}$
at $T \sim T^*=0.56T_{\rm c}$.
The meeting of the two phase boundaries
at $T^*=0.56T_{\rm c}$ is a robust feature
of models of the FFLO,
irrespective of dimensionality~\cite{shimprb}.
Note that the effect is very reproducible;
data for different samples and different cooling
and bias conditions are shown in Fig.~\ref{fig2};
all follow the same trends~\cite{janeloff}.

In summary, when the field lies in the quasi-two-dimensional
planes of $\kappa$-(BEDT-TTF)$_{2}$Cu(NCS)$_{2}$, there is
good evidence for a transition
into a Fulde-Ferrell-Larkin-Ovchinnikov (FFLO) state.
The data are in agreement with theoretical predictions~\cite{janeloff}.
\subsection{Field-induced superconductivity in $\lambda-$(BETS)$_2$FeCl$_4$}
\begin{figure}[htbp]
\centering
\includegraphics[height=12cm]{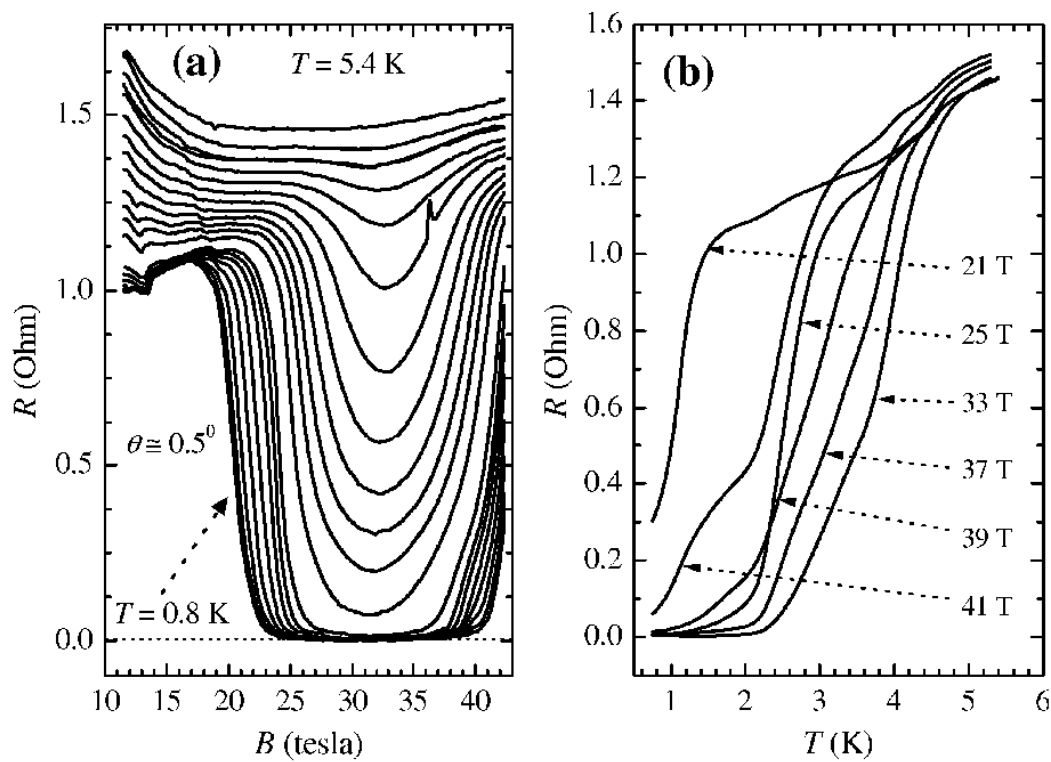}
\caption{Resistance of a crystal of $\lambda$-(BETS)$_2$FeCl$_4$
(a)~as a function of magnetic field for temperatures between
800~mk and 5.4~K and (b)~as a function of temperature for
various magnetic fields (after Reference~\cite{balicas}).
The magnetic field is applied within the quasi-two-dimensional planes of the crystal.
Note the strong minimum in the resistance, suggestive of
superconductivity.
}
\label{luis}
\end{figure}

There has been considerable interest in the material
$\lambda$-(BETS)$_2$FeCl$_4$, which shows
a transition to what appears to be a superconducting state
in an accurately in-plane magnetic field~\cite{uji} (Figure~\ref{luis}).
In this salt, the
Fe$^{3+}$ ions within the FeCl$_4$ anion molecules
order antiferromagnetically
at low temperatures and low magnetic fields (see Reference~\cite{uji}
and references therein).
As is the case in most antiferromagnetic materials,
the magnetic order causes $\lambda$-(BETS)$_2$FeCl$_4$ to become an insulator.
On raising the magnetic field (applied exactly within the planes),
the magnetic moments  of the Fe$^{3+}$ ions
tilt (i.e. become
``canted''), and then the long-range magnetic order is destroyed,
leaving a paramagnetic metal~\cite{uji}. Further increases in field
induce the superconducting state at about 16~T~\cite{uji}.

Field-induced superconductivity in magnetic materials
is usually discussed in terms of the Jaccarino--Peter compensation (JPC) effect,
in which the applied field ``compensates'' the internal
magnetic field provided by the magnetic ions (in the
case of $\lambda$-(BETS)$_2$FeCl$_4$, the Fe$^{3+}$ ions).
The JPC effect was predicted in the early 1960s~\cite{jacpet},
and observed almost
twenty years ago in Eu$_x$Sn$_{1-x}$Mo$_6$S$_8$~\cite{oldstuff}.

Therefore, the situation in $\lambda$-(BETS)$_2$FeCl$_4$
seems to be as follows. The applied magnetic field compensates
the internal field provided by the Fe$^{3+}$ ions,
so that superconductivity (which would otherwise be suppressed by the magnetism)
can occur at fields above 16-20~T~\cite{uji,balicas}.
The previous sections have shown that
a quasi-two-dimensional superconductor can attain high upper critical fields
{\it if the magnetic field is applied exactly in the
plane of the layers}. In such a configuration, only a very small
number of closed orbits can occur on the Fermi surface.
Hence, the orbital interactions which would otherwise
overwhelm the superconductivity are suppressed~\cite{msnsuper}.
The magnetic field can then only destroy the superconductivity
via the Zeeman effect, leading to an upper critical
field determined by the Pauli limit or
a more exotic mechanism such as the FFLO.
It is notable that the field-induced superconductivity in
$\lambda$-(BETS)$_2$FeCl$_4$ {\it only} occurs when the
magnetic field is exactly within the layers.

Because of this, with the magnetic field applied exactly within the layers,
$\lambda$-(BETS)$_2$FeCl$_4$, in common with other quasi-two-dimensional
organic superconductors (see Sections~\ref{s4p2} and \ref{s5p1}),
has a high upper critical field
because orbital effects are suppressed.
This allows the material
to continue to superconduct to fields well above 16~T.
Recent measurements~\cite{balicas}
suggest that the upper critical field
is around 40~T,
the material regaining measurable resistance above this field (Figure~\ref{luis}).

Whilst this simple picture accounts qualitatively for most aspects
of the data presented thus far, there are undoubtedly
further complications, and much work remains to be done.
Whereas the Cooper pairs in
conventional superconductors are formed via electron-phonon
interactions, there is evidence that the pairing interactions in organic
superconductors involve magnetic excitations (see Section~\ref{s3p1p1}).
The interaction between these excitations and the
magnetic moments of the Fe$^{3+}$ ions
undoubtedly complicates the simple effects described above.
\subsection{Persistent currents in the high-field state of \khg}
Prior to the development of the BCS theory,
Fr\"{o}hlich proposed that superconductivity would occur in
quasi-one-dimensional metals due to spontaneously sliding
charge-density waves (CDWs)~\cite{oldfrohlich}. CDWs were subsequently
discovered in many materials with quasi-one-dimensional Fermi-surface
sections~\cite{gruner}. However,
the opening of the energy gap, and pinning to impurities,
generally prevent a CDW from contributing to the
electrical conductivity, so that the transport properties
are dominated by any remaining normal carriers~\cite{gruner}.
Depinning of CDWs may occur
under large electric fields, but is accompanied by considerable
dissipation.

However, recently, Harrison et al. have made measurements that
appear to support the presence of something like
a superconducting state in \khg ~at high magnetic fields.
This may be a a Fr\"{o}hlich superconductor, or
could possibly entail a more exotic mechanism involving
the dynamic exchange of quasiparticles between the
density wave in the quasi-one-dimensional Fermi-surface
section and a quasi-two-dimensional pocket
also present~\cite{science}.

\begin{figure}[htbp]
\centering
\includegraphics[height=12cm]{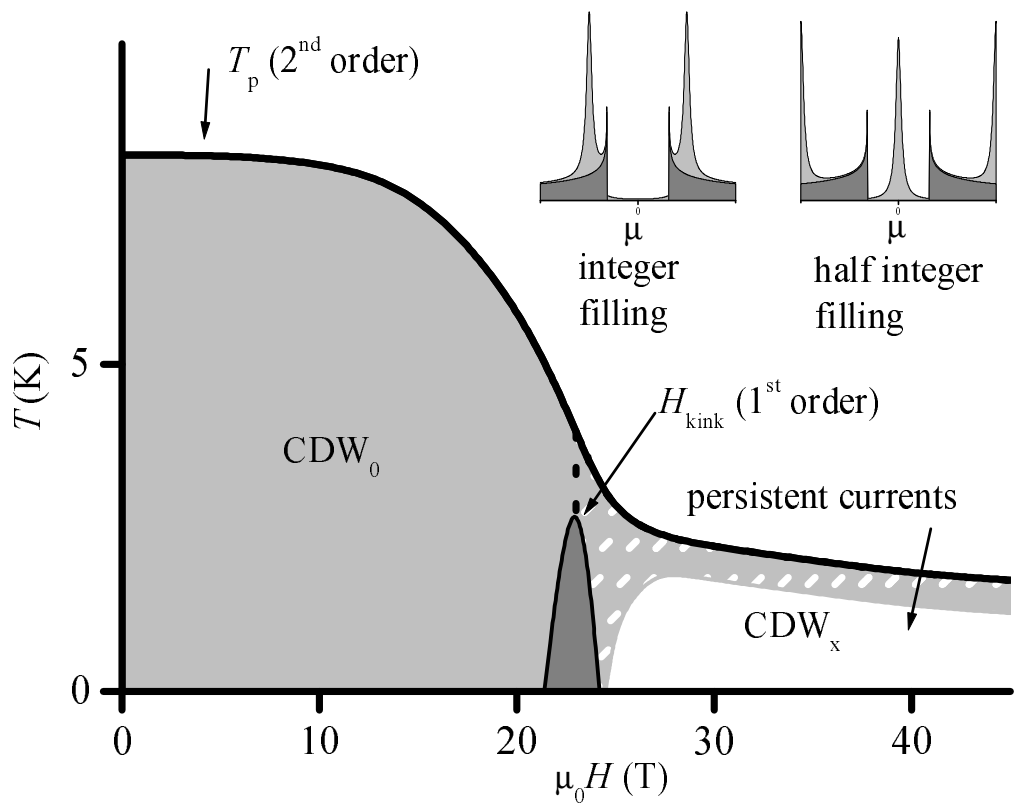}
\caption{Notional phase diagram of $\alpha$-(BEDT-TTF)$_2$KHg(SCN)$_4$
constructed from theoretical models and data
(see Reference~\cite{frohlich}). The solid
line represents a second order transition into the CDW phase (light shading) with
a dotted line [together with the region of hysteresis (heavy shading)]
representing a first order transition between the proposed CDW$_0$
(solid shading) and CDW$_x$ (hatched shading) phases. Persistent
currents (white region) are observed only within the CDW$_x$
phase. The top right-hand corner shows the density of states
resulting from CDW formation within the 1D bands (heavy shading) together
with the contributions from the Landau levels of the two dimensional band (light
shading), both at integer and half integer filling factors. $\mu$
represents the position of the chemical potential.
}
\label{diagram}
\end{figure}

In its normal state, \khg ~has a Fermi surface consisting of a
quasi-two-dimensional
pocket and a pair of sheets, the sheets nesting at low temperatures
to give the commensurate CDW$_0$ phase~\cite{review,frohlich}
The phase diagram is shown in Figure~\ref{diagram};
at a field of 23~T, the CDW$_0$ phase
reaches its Pauli paramagnetic limit and transforms into
the incommensurate CDW$_x$ phase, with a very low transition
temperature $\sim 2$~K.
$\alpha$-(BEDT-TTF)$_2$KHg(SCN)$_4$ is therefore unusual in two
respects: (1) it exhibits a CDW state in which
the gap energy $\Delta_0$ is sufficiently small for such a transition to occur in
accessible magnetic fields; and (2) the
reduced gap
$2\Delta_x\approx$~1~meV that characterizes the CDW$_x$ phase
is $\sim 10^2$ smaller than those in
typical CDW materials~\cite{gruner}.

In the Fr\"{o}hlich interpretation of Harrison's data,
the smallness of the gap
means that the zero-point energy of the CDW$_x$ phase dramatically
exceeds the pinning potential, allowing the CDW to slide freely~\cite{frohlich}.
In the alternative explanation~\cite{science},
the smallness of the pinning potential
is also a key factor, allowing the CDW to change
its nesting vector
(with some resistance)
in response to the changing field.
Whatever the explanation,
within the CDW$_x$ phase (see Figure~\ref{diagram}),
robust persistent currents are seen in magnetisation data,
behaving in exactly the same manner as do those found in
an inhomogeneous type II superconductor~\cite{frohlich}.

Interestingly, the persistent currents seen in \khg
~were for some time attributed to the quantum Hall effect
and its attendant chiral Fermi liquid
(see Section 4.6.2 of Reference~\cite{review} and
Reference~\cite{honold} for a review).
The CDW formation is of course due to the
quasi-one-dimensional section of the Fermi surface;
however, in the CDW$_x$ phase, the density wave coexists
with very sharp Landau levels of the quasi-two-dimensional Fermi surface
(see inset to Figure~\ref{diagram}).
The presence of the Landau levels appears to modulate
the size and nature of the persistent
currents provided by the density wave~\cite{frohlich}.
It was this apparent Landau-level filling factor dependence
which led to proposals of the involvement of
the quantum Hall effect~\cite{frohlich,science}.
\section{Summary}
It is hoped that this review has outlined some of the
many reasons for studying quasi-two-dimensional
organic superconductors. These systems form a unique
``test-bed'' for ideas about the superconducting state;
in contrast to many other types of superconductor, their
(usually simple) Fermi surfaces may be readily measured to great accuracy.
Subsequently ``chemical'' or real pressure can be used
to vary the bandstructure controllably, and the consequent changes in
superconducting properties studied.

In spite of the strong interactions, a Fermi-liquid approach
seems at present
able to describe the properties of the normal-state quasiparticles
(see {\it e.g.}~\cite{msnloc}), although intensive searches continue
for for departures from this comforting behaviour.
However, much work remains to be done on the question
of the interactions involved in the superconductivity.
Although it is clear that electron-phonon
and electron-electron interactions, and
antiferromagnetic fluctuations all play a part
in the superconducting mechanism, their exact role
is yet to be determined.
Questions such as this, plus the emergence
of new organic superconductors with even more exotic phase diagrams~\cite{mck22}
will ensure that this field remains a very active one for some time to come.

\section{Acknowledgements}
Work on organic conductors at Los Alamos
is supported by the Department of Energy, the National
Science Foundation (NSF) and the State of Florida,
and that at Oxford is supported by
EPSRC (UK). We should
like to thank Albert Migliori, Paul Goddard,
Arzhang Ardavan, Anne-Katrin Klehe,
Francis Pratt, Stan Tozer, Neil Harrison,
Bill Hayes, Francis Pratt, Jim Brooks, Joerg Schmalian, Luis Balicas,
Phil Anderson and Steve Blundell
for useful discussions and encouragement.
Part of this article was written at the Aspen Center for Physics;
JS expresses his gratitude for the opportunity to work in
such a stimulating environment.

\section{Note added in proof}
Further strong evidence for d-wave superconductivity
in $\kappa$-phase BEDT-TTF salts comes from very
recent thermal conductivity experiments;
see K. Izawa, H. Yamaguchi, T. Sasaki and Y. Matsuda,
Phys. Rev. Lett. {\bf 88}, 027002 (2002).

\section{Biographical details}
After a doctorate in semiconductor physics,
and a postdoctoral research fellowship at Oxford University,
John Singleton worked at the High Field Magnet Laboratory,
Katholieke Universiteit Nijmegen.
It was in Nijmegen, on the advice
of Bill Hayes and Francis Pratt (Oxford), that
he began to study organic superconductors using high
magnetic fields.
In 1990, John returned to Oxford University
as a Lecturer in Physics (since promoted to Reader
in Physics). Since then he has helped to establish
the {\it Oxford Correlated Electron Systems Group},
a collection of about twenty academics, postdocs,
students and visitors working on organic metals
and magnets,
electrically-conductive metal oxides, heavy-fermion
compounds, small biological molecules and novel
solid-state light sources.
He spent the academic year 2000-2001 as a visiting
scientist at the National High Magnetic Field Laboratory,
Los Alamos.

Charles (Chuck) Mielke received his doctorate
(on various organic conductors) from
Clark University, Worcester, Massachusetts,
where he also helped to set up
the pulsed-field-magnet facility.
Whilst at Clark, Chuck played a pivotal role in
developing tunnel diode oscillators for
measuring the MHz skin depth of small superconducting
samples in
difficult environments (e.g. pulsed fields,
high pressures). He moved to Los Alamos National
Laboratory in 1996, and now works on a variety of
research projects involving
pulsed magnetic fields, high frequencies, power engineering,
cuprate and organic superconductors and heavy-fermion compounds.



\begin{thebibliography}{9}
\bibitem{footnote}
The existence of such a modulation might suggest that a beating
between ``neck and belly'' frequencies would be observed in the
dHvA effect~\cite{review}; however, in careful low-field studies~\cite{sasaki},
no such beating has been observed.
We shall see that this is because the typical cyclotron energy
is rather greater than $t_{\perp}$, evan at the lowest fields used.
\bibitem{cokebottle}
When viewed in the extended zone scheme, this type of Fermi surface
is often known as a ``coke-bottle'' Fermi surface (USA) or
a ``Cumberland sausage'' Fermi surface (Europe). (The latter
is a hand-made sausage with weak but approximately regular
corrugations.)
\bibitem{crdisclaimer}
Note that the cyclotron resonance data cited
in Reference~\cite{caulfield} are probably
erroneous~\cite{review}.
\bibitem{ishigutsq}
Some representative resistivity data showing the $T^2$
dependence are collected together in Figure~5.27 of
Reference~\cite{ishiguro}.
\bibitem{ishiguhump}
Earlier arguments concerning the nature of
the ``hump'' in the resistivity of the
$\kappa$-phase BEDT-TTF salts are collected
on page 151 of Reference~\cite{ishiguro}.
Some more recent views are described
in References~\cite{wang,tanatar}.
\bibitem{anguish}
The phase diagram of $\kappa$-(BEDT-TTF)$_2$Cu[N(CN)$_2$]Cl
shown at the left-hand side of Figure~\ref{nature}(b)
is considerably simplified.
The magnetic state of this salt shows
complicated behaviour as a function of temperature,
pressure and magnetic field,
as has been described in References~\cite{sinojap}.
\bibitem{estimate}
Taking the value of $\gamma \sim 100-360$
from Reference~\cite{Mansky},
the estimated in-plane critical field
if orbital effects were the only limiting factor
would be $B_{\rm c2}(\theta=90^{\circ}) =
\gamma B_{\rm c2}(\theta =0) \sim 500$~T at 1.45~K!
This is plainly far higher than the experimental value.
\end{thebibliography}
\end{document}